\documentclass{elsarticle}
\usepackage{amsmath,amssymb,a4wide,psfrag}
\newcommand{\Tr}[0]{\text{Tr}\;}
\newcommand{\sign}[0]{\text{sign}\;}
\newcommand{\FT}[1]{F_T\left\{#1 \right\}}

\begin{document}
\begin{frontmatter}
\title{A renormalisation group derivation of the overlap formulation}
\author[a]{Nigel Cundy}
\address[a]{Institut f\"ur Theoretische Physik, Universit\"at Regensburg, D-93040 Regensburg, Germany}

\begin{abstract}
Starting from the continuum Dirac operator, I construct a renormalisation group blocking which transforms the continuum action into a lattice action, and I specifically consider the Wilson and overlap formalisms. For Wilson fermions the inverse blocking is non-local and thus invalid. However, I proceed to demonstrate that it is possible to construct a valid, local, blocking which, though dependent on the lattice spacing, generates the lattice overlap fermion action from the continuum action. 

Using this renormalisation group blocking for overlap fermions, I re-derive the Ginsparg-Wilson equations and the lattice chiral symmetry, and show that the standard Ginsparg-Wilson relation is not the most general way of expressing chiral symmetry on the lattice, nor, for overlap fermions, the most natural. I suggest how this reformulation of the Ginsparg-Wilson relation combined with the renormalisation group formulation of overlap fermions could allow the construction of a $\mathcal{CP}$ invariant lattice chiral gauge theory.
\end{abstract}
\end{frontmatter}

\section{Introduction}
Chiral symmetry is one of the most important properties of the massless continuum QCD Lagrangian. In lattice QCD, however, it causes something of a problem.

The infinitesimal chiral transformation is usually given as
\begin{align}
\psi &\rightarrow \psi + i\upsilon\gamma_5 \psi\nonumber\\
\overline{\psi}&\rightarrow \overline{\psi} + i \upsilon \overline{\psi}\gamma_5,
\end{align}
and it is clear that the massless fermion action, with Dirac operator $D_0$, $\overline{\psi} D_0 \psi$, is invariant under this symmetry because 
\begin{gather}
\{D_0,\gamma_5\}=0.\label{eq:1cs}
\end{gather}
However, on the lattice, Nielsen and Ninomiya  showed that it is impossible to simultaneously satisfy equation (\ref{eq:1cs}) while maintaining translation invariance, locality, and having a theory without doublers~\cite{Nishy-Ninny}. An alternative way of expressing their no-go theorem is to say that any lattice theory must have an equal number of left- and right-handed fermions, which seems to forbid the existence of zero modes which in the continuum cause an imbalance in this number. Shortly afterwards, based on a construction derived from the renormalisation group, Ginsparg and Wilson described a way in which chiral symmetry could be maintained on the lattice, namely that the right-hand side of equation (\ref{eq:1cs}) could be modified to  give a term which is both local and which vanishes in the continuum limit~\cite{Ginsparg:1982bj}. However, no solutions were found, the Ginsparg-Wilson equation forgotten, and for over ten years the lattice community continued believing that chiral symmetry and the lattice were incompatible.

In the 1990s, three lattice Dirac operators were proposed which do satisfy Ginsparg and Wilson's equation, in two cases approximately (in practical simulations) and in the other exactly (up to working numerical precision). Kaplan noted that by switching to a 5-dimensional lattice, and treating the four-dimensional lattice as one wall of the 5D lattice, he could separate the left and right handed fermions by a large enough distance in the fifth dimension  that they would not interact~\cite{Kaplan:1992bt,Shamir:1993zy,Furman:1994ky}. By sending the size of the 5th dimension to infinity, this would give a chiral lattice Dirac operator, the domain wall fermion. In practice, the size of the 5th dimension cannot be increased to such a degree that the chiral effects can be utterly neglected, and domain wall fermions are only approximately chiral, albeit to an exceptionally good approximation.

Shortly afterwards, and inspired by Kaplan's work, Neuberger reasoned that if the Dirac operator described an infinite number of fermion fields, then one could also have a number of zero modes and the same number (i.e. $\infty$) of left- and right-handed fermions~\cite{Narayanan:1993sk,Narayanan:1993ss,Neuberger:1998fp,Neuberger:1998my}. This lead him to the overlap formula. The domain wall action reduces to a form of the overlap action at infinite fifth dimension. 

About the same time, several researchers were experimenting with the idea of applying a renormalisation group blocking to a gauge field on a coarse lattice to obtain another at a finer lattice~\cite{Hasenfratz:1994}, and this idea was later extended to incorporate lattice fermions~\cite{Bietenholz:1995nk,Bietenholz:1995cy}, and has since been suggested as a way to include super-symmetry on the lattice~\cite{Bergner:2008ws}. In principle, if the form of the lattice action would be invariant under such a blocking, then it would be classically perfect: a non-perturbative approximation to a perfect action, with small scaling artifacts. This could be achieved by repeating the blocking numerous times, starting with a suitable lattice operator, and the action would flow towards a fixed point, which would satisfy the lattice chiral symmetry. Unfortunately, it is not possible to implement a closed form of the fixed point action, so the blocking procedure has to be truncated, again leading to an imperfect chiral symmetry.

In the context of his work, Peter Hasenfratz rediscovered the Ginsparg-Wilson equation~\cite{Hasenfratz:1998ri}, and showed that his classically perfect fermions satisfied the lattice chiral symmetry. It was subsequently realised that overlap fermions also obeyed the Ginsparg-Wilson symmetry, and by extension domain wall fermions almost obey it, Martin L\"uscher discovered that the Ginsparg Wilson equation implied a symmetry of the lattice fermion action~\cite{Luscher:1998pqa}, and the age of lattice chiral fermions was born.

Subsequently, there have been numerous other either approximate~\cite{Bietenholz:1998ut,Gattringer:2000js,Gattringer:2000qu,Borici:1999zw} or exact~\cite{Cundy:2008cs,Fujikawa:2000my,Kerler:2002xk} chiral lattice Dirac operators proposed, although none offer any significant improvements over the original methods.

However, there are two outstanding theoretical issues concerning lattice chiral fermions which remain troubling. Firstly, the construction of a lattice chiral gauge theory using the standard Ginsparg-Wilson formalism violates $\mathcal{CP}$~\cite{Hasenfratz:2001bz,Fujikawa:2002is,Fujikawa:2002vj}, although, since $\mathcal{CP}$ is restored in the continuum limit, it is natural to assume that the effects of this violation are as negligible as the broken Lorentz symmetry on the lattice. It has recently been shown that this broken $\mathcal{CP}$ is related to an observation that the lattice Ginsparg-Wilson Dirac operators do not just obey one chiral symmetry but an infinite group of chiral symmetries, each with a different (unrenormalised) current~\cite{Mandula:2007jt,Mandula:2009yd}.

 Secondly, and troubling at more of a theoretical than practical level, the Ginsparg-Wilson equation and the fixed point fermions were derived from renormalisation group considerations; while overlap and domain wall fermions were derived by an entirely different approach. That they satisfy (or approximately satisfy) the Ginsparg-Wilson relation hints that there could be some relationship between these operators and the renormalisation group. An understanding of this relationship would tie up a loose end to the theoretical basis of lattice chiral symmetry. It could also, in principle (if not necessarily in practice), be used to calculate the renormalisation group coefficients of the action, or to take a continuum limit without an extrapolation (in the lattice spacing), since if the lattice theory were linked to the continuum theory by a renormalisation group transformation, then the continuum limit could be achieved just by calculating the appropriate renormalisation constants.

Until this work, there was no known relationship between the overlap operator and the renormalisation group. Here, I shall derive the (infinite volume) overlap action by applying a simple, exponentially local, renormalisation group blocking to the continuum Euclidean fermion action.
 I shall demonstrate by extensive calculation that this blocking is valid, in the sense that it is analytic, local and reversible. A corollary is that it is possible to apply a blocking to the overlap action that gives the continuum action without taking any zero lattice spacing limit. Thus, (under certain conditions) overlap lattice  QCD is not just a theory that approaches continuum (Euclidean) QCD, it is continuum QCD in a particular (and somewhat peculiar) renormalisation scheme.

This formalism naturally leads to a discussion of the construction of a chiral gauge theory on the lattice. Most attempts at the construction of a lattice chiral gauge theory using Ginsparg-Wilson fermions have failed  because any projection operators which obey $\mathcal{CP}$ symmetry must have singularities in the Brillouin zone~\cite{Fujikawa:2002is,Jahn:2002kg}\footnote{A possible solution to this problem, which uses a lattice redefinition of the Parity operator, a different approach to that presented here, was recently suggested in~\cite{Igarashi:2009kj}; see also the attempts using the perfect action formalism in ~\cite{Hasenfratz:2007dp,Gattringer:2007dz}. An older overview of chiral gauge theories on the lattice can be found in~\cite{Golterman:2000hr}.}. However, this argument was only constructed using the standard form of the Ginsparg-Wilson equation; and I will suggest that $\mathcal{CP}$-invariant chiral gauge theories are possible on the lattice with a different formulation of the Ginsparg-Wilson relation.

In section \ref{sec:2}, I describe the theory behind block renormalisation transformations, and, in section \ref{sec:3}, I review the Ginsparg-Wilson relation and associated chiral symmetry. In section \ref{sec:4}, I construct an (invalid) blocking which, were it valid, would allow Wilson fermions to be derived from the continuum operator. I use the results of this section to construct the overlap operator and standard overlap chiral symmetry in section \ref{sec:5}, and, in section \ref{sec:7}, I use a different blocking to generate a symmetric Ginsparg-Wilson equation, which I use, in section \ref{sec:8}, to construct a possible $\mathcal{CP}$-invariant lattice chiral gauge theory, and I discuss this theory further in section \ref{sec:8b}. I divulge some concluding remarks in section \ref{sec:9}. There are appendices giving my notation and the proofs of a few results which are needed in the text.

A preliminary outline of sections \ref{sec:2}-\ref{sec:7} was presented in reference~\cite{Cundy:2008cn}.
\section{Block renormalisation group transformations}\label{sec:2}
I define a Block renormalisation group transformation from a fermion field $\psi_0$ with Dirac operator $D_0$ to a fermion field $\psi_1$ with a Dirac operator $D_1$ in terms of three functions $\hat{B}$, $\hat{\overline{B}}$ and $\alpha$:
\begin{align}
Z_0=&N(\alpha)\int d\psi_0d\overline{\psi}_0 e^{-\overline{\psi}_0D_0\psi_0 -\frac{1}{4g_0^2}F^2}\int d\psi_1 d\overline{\psi}_1e^{-(\overline{\psi}_1 - \overline{\psi}_0 \hat{\overline{B}})\alpha(\psi_1 - \hat{B}\psi_0)}\nonumber\\=& N'(\alpha)\int d\psi_1d\overline{\psi}_1 e^{-\overline{\psi}_1 D_1 \psi_1 -\frac{1}{4g_1^2}F^2},\label{eq:1}
\end{align}
where $N$ and $N'$ are normalisation constants, 
and $F$ is the field strength tensor. I stress that I am only blocking the fermion fields: the continuum gauge field $A_{\mu}$ is retained throughout this work. It is usual when considering renormalisation group blocking transformations of the type described in equation (\ref{eq:1}) to block from one manifold to another, so, for example, the fermion field $\psi_1$ could be on the lattice while $\psi_0$ on the continuum, or they could be spinor fields on lattices with two different lattice spacings. However, in this work, I take a different approach: $\psi_1$ and $\psi_0$ will be different representations of continuum spinor fields, with the action $\overline{\psi}_1D_1\psi_1$ only reducing to a lattice action in a particular limit, which will be taken at the end of the calculation; and even then, although the action will be identical to the lattice action, the spinor fields will (formally) remain continuum spinor fields. To preserve gauge covariance, $\hat{B}$ and $\hat{\overline{B}}$ must be functions of the gauge field, while $\alpha$ must be independent of the gauge field  so that the normalisation constant N($\alpha$) commutes with the (suppressed in equation (\ref{eq:1})) integration over the gauge field $A_{\mu}$. Both $B$ and $\alpha$ may contain a non-trivial Dirac structure.  Because this is Euclidean space-time, there is no need for $\hat{\overline{B}}$ and $\hat{B}$ to be conjugate, since $\psi$ and $\overline{\psi}$ are treated as independent variables; and in general I shall treat $\hat{B}$ and $\hat{\overline{B}}$ as independent. This may cause difficulties when analytically continuing to Minkowski space-time, and it will be important to take appropriate limits before calculating physical results. Throughout this article, I shall use one flavour of massless fermions, but the extension to multiple flavours is straightforward.
For the moment, in this general discussion, I assume that $\hat{B}$, $\hat{\overline{B}}$ and $\alpha$ are all invertible, leaving the proof for specific examples to a later section. I shall only work in the continuum (although with a theory equivalent to that found on the lattice as my aim), thus $\hat{B}$, $\hat{\overline{B}}$ and $\alpha$ are all square rather than rectangular matrices\footnote{The reader should bear in mind that these objects are not in reality `matrices' but linear operators of functions, even though I shall use the conceptionally easier language of matrices throughout this article.}. I will also assume that there are no complications when taking the infinite volume limit. It can easily be shown that
\begin{align}
Z_0 =& \int d\psi_0 d\overline{\psi}_0 d\psi_1 d\overline{\psi}_1e^{-\frac{1}{4g_0^2}F^2}e^{-\overline{\psi}_1(\alpha - \alpha \hat{B}\frac{1}{D_0 + \hat{\overline{B}}\alpha \hat{B}}\hat{\overline{B}}\alpha)\psi_1}\nonumber\\
& e^{-(\overline{\psi}_0 - \overline{\psi}_1\alpha \hat{B}\frac{1}{D_0+\hat{\overline{B}}\alpha \hat{B}})(D_0+\hat{\overline{B}}\alpha \hat{B})(\psi_0 - \frac{1}{D_0+\hat{\overline{B}}\alpha \hat{B}} \hat{\overline{B}}\alpha \psi_1)}.
\end{align}
I set $\alpha$ to be proportional to the unit matrix, and take the limit as $\alpha \rightarrow \infty$, while assuming that $\hat{B}$ is just a function of the $\gamma$ matrices and the gauge fields. Shifting the variables 
\begin{align}
\psi_0 \leftarrow& \psi_0 - \frac{1}{D_0+\hat{\overline{B}}\alpha \hat{B}} \hat{\overline{B}}\alpha \psi_1 ,\nonumber\\
\overline{\psi}_0 \leftarrow &\overline{\psi}_0 -  \overline{\psi}_1\alpha \hat{B}\frac{1}{D_0+\hat{\overline{B}}\alpha \hat{B}},
\end{align}
and using
\begin{align}
\alpha - \alpha \hat{B}&\frac{1}{D_0 + \hat{\overline{B}}\alpha \hat{B}}\hat{\overline{B}}\alpha =\nonumber\\&
 -\alpha \hat{B} \frac{1}{\hat{\overline{B}}\alpha \hat{B}} D_0  \frac{1}{\hat{\overline{B}}\alpha \hat{B}}\hat{\overline{B}}\alpha + O(\alpha^{-1})
\end{align}
allows the integration over the shifted $\psi_0$ fields if the spectrum of $D_0+\hat{\overline{B}}\alpha \hat{B}$ contains only eigenvalues whose real part is greater than zero. Once again, this condition will have to be tested for specific examples. The result of this integration is
\begin{gather}
Z_1 = \int d\psi d\overline{\psi} e^{-\overline{\psi}_1\overline{B}D_0 B \psi_1}e^{-\frac{1}{4g_0^2}F^2}e^{\Tr\log(\hat{\overline{B}} \hat{B})},
\end{gather}
where I have defined
\begin{align}
\alpha \hat{B} \frac{1}{\hat{\overline{B}}\alpha \hat{B}} =& \overline{B},\nonumber\\
 \frac{1}{\hat{\overline{B}}\alpha \hat{B}}\hat{\overline{B}}\alpha =& {B}.\label{eq:lotsofBs}
\end{align}
This is satisfied if
\begin{align}
B =& \hat{B}^{-1},& \overline{B} =& \hat{\overline{B}}^{-1},\label{eq:thebleedingobvious}
\end{align}
and I shall use this less general definition throughout this work, and shall describe $\hat{B}$ and $\hat{\overline{B}}$ as the inverse of the blockings. However, it should be noted that there may be occasions when $1/(\hat{\overline{B}}\alpha \hat{B})$ is well defined while $1/\hat{B}$ is not (for example, if the matrices $B$ and $\overline{B}$ were rectangular), when it would be necessary to generalise and use the definition contained within equation (\ref{eq:lotsofBs}) rather than equation (\ref{eq:thebleedingobvious}). 
I will call $\Tr\log(\hat{\overline{B}} \hat{B})$ the Jacobian of the blocking.

The new fermion action is $\overline{\psi}_1D_1\psi_1$ where $D_1 = \overline{B}D_0 B$. It should be noted that $B$ and $\overline{B}$ are not the only transformations that can be used to derive a Dirac operator $D_1$. For example (and these are not the only examples), given a suitable transformation, and any invertible operator $A_1$ which commutes with $D_1$ and any invertible  $A_0$ which commutes with $D_0$, another possible set of blockings is given by 
\begin{align}
B' =& A_0 B A_1,\nonumber\\
\overline{B}' =& A_1^{-1} \overline{B}A_0^{-1}\label{eq:degenerateblockings}.
\end{align}
I shall use this degeneracy in section \ref{sec:7}.

Thus this is a valid renormalisation group transformation if the following conditions are satisfied:
\begin{enumerate}
\item Given a $\hat{B}$ and $\hat{\overline{B}}$ there exist a $B$ and $\overline{B}$ according to equation (\ref{eq:thebleedingobvious}) (or the more general form in equation (\ref{eq:lotsofBs}));
\item $\hat{\overline{B}}$, $\hat{B}$, $B$ and $\overline{B}$ are all local (in the sense that $B(x,y) \le e^{-a^{-1}|x-y|}$ for some small $a$);
\item $D+\hat{\overline{B}}\alpha \hat{B}$ has no eigenvalues with negative or zero real part;
\item The blocking is gauge-covariant;
\item The Jacobian reduces to a constant, Yang-Mills term and (possibly) some irrelevant operators.
\end{enumerate}
The final point is straightforward to prove, and holds for all possible exponentially local blockings.
If $\log(\hat{\overline{B}} \hat{B})$ is a function of only the gauge fields and the $\gamma$-matrices, then $\Tr\log(\hat{\overline{B}} \hat{B})$ must consist of a constant term and closed loops of the gauge fields, so that
\begin{gather}
\Tr\log(\hat{\overline{B}} \hat{B}) = \int d^4x \sum_{C[x]} w(C[x]) P\left[e^{ig\int_{C[x]} A_{\mu}(x')dx'}\right],
\end{gather}
where $C[x]$ is a closed loop starting and ending at $x$, $w(C)$ is a weight function (possibly a function of the $\gamma-$matrices), and $P$ represents path ordering. If both $D_0$ and $D_1$ are $\gamma_5-$Hermitian, then there will be at least one possible choice of blockings where $B^{\dagger} = \gamma_5\overline{B}\gamma_5$, namely $B=D_0^{-1/2}D_1^{1/2}$, $\overline{B} = D_1^{1/2}D_0^{-1/2}$. Thus this particular $\hat{\overline{B}}\hat{B}$ is $\gamma_5$-Hermitian. From the cyclicity of the trace, the Jacobian must be the same for every possible blocking generated according to equation (\ref{eq:degenerateblockings}). This means that any anti-Hermitian component of $\hat{\overline{B}} \hat{B}$ (or any function of it) must be proportional to $\gamma_{\mu}$ and will be traceless. Therefore if I expand $\Tr\log \hat{\overline{B}}\hat{B}$ in terms of the gauge fields and $\gamma$-matrices, only Hermitian terms can survive. The trace must be composed of closed loops of gauge links, which restricts it to terms constructed from the anti-Hermitian field strength tensor $F$.
Therefore the Jacobian must be the sum of operators constructed from the field strength tensor and its derivatives: $\log(\hat{\overline{B}} \hat{B}) = c_0 + c_1\sigma_{\mu\nu}F_{\mu\nu} + c^{(1)}_2F_{\mu\nu}^2 + c_2^{(2)}F_{\mu\nu}\tilde{F}_{\mu\nu}  +  \ldots$. The constant term will not contribute to any physics, and can be neglected. The $\sigma F$ term is traceless. The $F\tilde{F}$ term is forbidden if both $\overline{\psi}_1D_1\psi_1$ and $\overline{\psi}_0D_0\psi_0$ are invariant under $\mathcal{CP}$ (see appendix \ref{app:CP1}).  Thus in such an expansion, the dominant term will be $F_{\mu\nu}^2$, with higher powers of $F$ suppressed by $a$, the range of the locality of the blocking operators $B$ (this follows from dimensional analysis, given that $a$ is the only quantity with dimensions of length available).
  Hence, if $B$ is sufficiently local that the higher order terms can be neglected, this Jacobian is proportional to the Yang-Mills gauge action and just entails a change in the coupling constant.

Indeed, it has previously been shown that the `natural' (though expensive) way of simulating the Yang-Mills action on the lattice is through the trace of a function of the Dirac operator~\cite{Horvath:2006aj,Horvath:2006md,Liu:2007hq}.

Any blocking leading to a Dirac operator which has a different number of exact zero modes to the continuum operator fails these tests. This is true for any lattice fermion action except Ginsparg-Wilson operators (on sufficiently fine lattice spacing) in non-trivial topological sectors. Throughout this work I will assume without proof or discussion that the index of the lattice overlap operator matches the index of the continuum operator as long as the lattice spacing is sufficiently fine\footnote{This has been demonstrated for the finite volume torus~\cite{Adams:2000rn} and certain infinite volume settings~\cite{Adams:1998eg}, although in an infinite volume the issue is non-trivial and depends to an extent on the gauge field and choice of kernel, for example, see the discussion in~\cite{Chiu:2001bg}.}. Clearly, the conclusions of this work depend on the validity of this assumption.
This can be shown by considering the zero modes, $\phi_0$, of
$
D_0 = \hat{\overline{B}}D_1\hat{B}
$. If $D_0\phi_0 = 0$, then (given that $\overline{B} = B^{\dagger}$) either $\hat{B}\phi_0 = 0$ or $D_1 B \phi_0 = 0$; but for Wilson fermions (as an example) $D_1$ has no exact zero modes; hence $\hat{B}$ must have a zero mode and thus also $\hat{\overline{B}} \hat{B}$, which violates the condition that $D+\hat{\overline{B}}\alpha \hat{B}$ has no eigenvalues with negative or zero real part. Similarly, if the Dirac operator $D_1$ has more exact zero modes than the continuum operator, the blocking from the lattice to the continuum will be invalid. Also any lattice Dirac operator with doublers will be forbidden for the same reason.

One concern with this approach is that I am introducing Dirac operators without point-like locality in the continuum theory. In practice, the requirement that the old and new Dirac operators have the same number of zero modes, which only holds if the lattice spacing is sufficiently fine (and, of course, which lattice spacings are `sufficiently fine' for a given configuration is somewhat unclear), places a natural bound on the locality of the action: the lattice spacing, and thus the rate of the exponential decay of the Dirac operator, will have to be significantly smaller than the smallest instanton in the system\footnote{It follows, if instantons of infinitesimal size are possible but exceptionally improbable at a finite volume, that it is necessary to take the $a\rightarrow0$ limit before the $V\rightarrow\infty$ limit, because the resolution of the lattice Dirac operator depends on the lattice spacing. If they are possible and probable enough to be seen on a finite volume, then this procedure will not work if such small instantons are allowed.}. Since I am aiming for a lattice theory, it is inevitable that at some level I will have to violate the usual continuum point-like locality. 

\section{The Ginsparg-Wilson symmetry}\label{sec:3}
This section is a review and generalisation of the work of Ginsparg and Wilson~\cite{Ginsparg:1982bj} and Martin L\"uscher~\cite{Luscher:1998pqa}, and almost all of the results presented here have been derived previously, for example in ~\cite{Hasenfratz:2007dp,Gattringer:2007dz} and, most particularly, in~\cite{Borici:2007ft,Borici:2007bp}. I differ from almost all previous authors considering renormalisation group blockings within the context of lattice QCD (with the exception of~\cite{Borici:2007ft,Borici:2007bp}) because I do not assume, as they did, that the blocking matrices commute with $\gamma_5$, and because, whereas they blocked from a continuum theory to a lattice or a lattice to another lattice with a different lattice spacing, I will block from a continuum theory with one action to another continuum theory but with a different action.
\subsection{Mass regularisation}\label{sec:3.1}
Since I will later need to construct blockings containing terms such as $(D_0)^{-1}$, it is necessary to regularise $D_0$, and I will do so by introducing an infinitesimal twisted mass $D_0\rightarrow D_0 +i\gamma_5\eta$, where $\eta$ is real. This shifts the zero mode eigenvalues of $D_0$ to $\pm i \eta$ and the non-zero modes from $\pm i\lambda$ (where $\lambda$ is real) to $\pm i \sqrt{\lambda^2 + \eta^2}$, so the inverse Dirac operator is now well defined. Accordingly, the Fourier transform of the Green's function associated with the Dirac operator will contain terms of order $1/\eta$, but no worse. For example, in the free theory the eigenvectors of the Fourier transform of the inverse Dirac operator are $\pm 1/\sqrt{p^2 + \eta^2}$. At the end of the argument, I shall take the limit $\eta \rightarrow 0$ if the limit exists. This regularisation obviously breaks $\gamma_5$-Hermiticity and $\mathcal{CP}$-symmetry, but these are restored as $\eta\rightarrow 0$. It does, however, preserve chiral symmetry and the structure of the eigenvalues of the Dirac operator. The infinitesimal chiral symmetry transformation is modified to
\begin{align}
\overline{\psi} \rightarrow& \overline{\psi} + \overline{\psi}i\upsilon\left(\gamma_5 - \frac{i\eta}{D_0 + i\eta\gamma_5}\right),\nonumber\\
\psi\rightarrow&\psi + i \upsilon\left(\gamma_5 - \frac{i\eta}{D_0 + i\eta\gamma_5}\right)\psi,\label{eq:chiral_symmetry}
\end{align} 
where $\upsilon$ is some small real number. As $\eta\rightarrow 0$, this becomes an operator projecting the zero modes of the Dirac operator from $\overline{\psi}\gamma_5$. The mass regularisation commutes with $D^{\dagger}D$, so that the eigenvectors of the Dirac operator are affected only by mixing between the non-zero eigenvector pairs.
This transformation is no longer ultra-local, but it will be exponentially local, since the Fourier transform of $i\eta/(D_0 + i\eta\gamma_5)$ is analytic, even in the limit that $\eta \rightarrow 0$\footnote{This operator is only local in Euclidean space, but it is non-local in Minkowski space-time because the branch cut in the Fourier transform is transfered from the imaginary to the real axis for certain components of the momentum. This, of course, makes continuation to Minkowski space-time harder, and, as with the lattice theory in general, it is necessary to take the appropriate continuum limits before continuing to Minkowski space-time.}. I shall write\footnote{This shift in the generator of chiral symmetry is related to the (non-local) zero-mode shift symmetry of the Euclidean (massless) Lagrangian: $\psi \rightarrow \psi + \alpha |\psi_0\rangle\langle \psi_0|\psi$ and $\overline{\psi}\rightarrow \overline{\psi}+ \alpha\overline{\psi} |\psi_0\rangle\langle \psi_0|$, a transformation which is invalid in Minkowski space.  To reconstruct the topological charge, it is necessary to `undo' this eigenvalue shift by explicitly adding the zero mode contribution back into the fermion fields. I Note that, if $\gamma_5$ is expressed in terms of the basis of the eigenvectors of $D_0$, which I will later need, then it is not traceless and, instead, $\Gamma_5$ is traceless.} 
\begin{align}
\Gamma_5 =& \gamma_5 - \frac{i\eta}{D_0 + i\eta\gamma_5}.\label{eq:Gamma5}
\end{align}

\subsection{The Ginsparg-Wilson relation}\label{sec:3.2}
 Taking the block transformation
\begin{gather}
Z_1 = \int d\psi_1 d\overline{\psi}_1d\psi_0 d\overline{\psi}_0e^{-\overline{\psi}_0 (D_0 + i\epsilon\gamma_5) \psi_0} e^{-(\overline{\psi}_1 - \overline{\psi}_0\hat{\overline{B}})\alpha(\psi_1 - \hat{B}\psi_0)},
\end{gather}
suppose that the original action is invariant under the infinitesimal symmetry defined by equations (\ref{eq:chiral_symmetry}) and (\ref{eq:Gamma5}). 
Neglecting terms of order $\upsilon^2$ and higher, and demanding that the new action is also invariant under the equivalent transformation, gives
\begin{align}
0 =& i\upsilon\int d\psi_1 d\overline{\psi}_1d\psi_0 d\overline{\psi}_0e^{-\overline{\psi}_0 (D_0 + i\eta\gamma_5) \psi_0} e^{-(\overline{\psi}_1 - \overline{\psi}_0\hat{\overline{B}})\alpha(\psi_1 - \hat{B}\psi_0)}\nonumber\\
&\phantom{spacespacespace}\left(\overline{\psi}_0\Gamma_5\hat{\overline{B}}\alpha(\psi_1 - \hat{B} \psi_0) + (\overline{\psi}_1 - \overline{\psi}_0 \hat{\overline{B}})\alpha \hat{B}\Gamma_5\psi_0\right).
\end{align}
Using the relations
\begin{align}
(\overline{\psi}_0\hat{\overline{B}} - \overline{\psi}_1)\alpha &e^{-(\overline{\psi}_1 - \overline{\psi}_0\hat{\overline{B}})\alpha(\psi_1 - \hat{B}\psi_0)} = \frac{\partial}{\partial \psi_1} e^{-(\overline{\psi}_1 - \overline{\psi}_0\hat{\overline{B}})\alpha(\psi_1 - \hat{B}\psi_0)},\nonumber\\
\alpha(\hat{B}\psi_0 -\psi_1)&e^{-(\overline{\psi}_1 - \overline{\psi}_0\hat{\overline{B}})\alpha(\psi_1 - \hat{B}\psi_0)} =  \frac{\partial}{\partial \overline{\psi}_1} e^{-(\overline{\psi}_1 - \overline{\psi}_0\hat{\overline{B}})\alpha(\psi_1 - \hat{B}\psi_0)},
\end{align}
I obtain
\begin{align}
0=&\left[\left(\frac{\partial}{\partial\psi_1} \alpha^{-1} - \overline{\psi}_1\right)\overline{B}\;\Gamma_5\hat{\overline{B}}\frac{\partial}{\partial \overline{\psi}_1} 
+ \frac{\partial}{\partial {\psi}_1}\hat{B}\Gamma_5 B\left( \alpha^{-1}\frac{\partial}{\partial\overline{\psi}_1} - \psi_1\right) 
\right]\nonumber\\
&\phantom{spacespacespacespacespace}\int d\psi_0 d\overline{\psi}_0 e^{-(\overline{\psi}_1 - \overline{\psi}_0\hat{\overline{B}})\alpha(\psi_1 - \hat{B}\psi_0)}e^{-\overline{\psi}_0 (D_0 + i\epsilon\gamma_5) \psi_0},\label{eq:GW:1}
\end{align}
where the partial derivatives are understood to only act on the partition function.
 From equation (\ref{eq:GW:1}) and the definition of $D_1$ given in equation (\ref{eq:1}), I derive
\begin{align}
0 =&\int d\psi_1 d\overline{\psi}_1\overline{\psi}_1\left[D_1\alpha^{-1} \overline{B}\;\Gamma_5\hat{\overline{B}} D_1  + D_1\hat{B}\Gamma_5 B \alpha^{-1} D_1 -\phantom{a}\right.\nonumber\\
&\phantom{spacespacespacespacespace}\left.\phantom{spacespace} \overline{B}\;\Gamma_5\hat{\overline{B}} D_1 - D_1\hat{B}\Gamma_5 B\right]\psi_1e^{-\overline{\psi_1}D\psi_1}. 
\end{align}
It is now trivial to construct the Ginsparg-Wilson relation:
\begin{gather}
D_1\alpha^{-1} \overline{B}\;\Gamma_5\hat{\overline{B}} D_1  + D_1\hat{B}\Gamma_5 B \alpha^{-1} D_1 = \overline{B}\;\Gamma_5\hat{\overline{B}} D_1 + D_1\hat{B}\Gamma_5 B.\label{eq:GW}
\end{gather}
If $[B,\Gamma_5]=0$ and $[\overline{B},\Gamma_5]=0$ this reduces to Ginsparg and Wilson's original result. 
However, this more general form (which is not original to this work, see, for example, ~\cite{Cundy:2008cs,Borici:2007bp}) allows different expressions of chiral symmetry on the lattice, and is crucial for avoiding the various no go theorems concerning the construction of a $\mathcal{CP}$-invariant chiral gauge theory (see section \ref{sec:8}). 
\subsection{Chiral symmetry}
Now suppose that the fermion action $\overline{\psi}_1 D_1 \psi_1$ is invariant under a `chiral' rotation given by
\begin{align}
\psi_1\rightarrow &e^{i\upsilon(S-\Gamma_5 RD_1)}\psi_1,\nonumber\\
\overline{\psi}_1\rightarrow & \overline{\psi}_1e^{i\upsilon(\overline{S} - D_1 \overline{R}\;\Gamma_5)}.
\end{align}
In the infinitesimal limit, the action transforms as
\begin{gather}
\overline{\psi}_1 D_1 \psi_1 \rightarrow \overline{\psi}_1 D_1 \psi_1 + i \upsilon \overline{\psi}_1\left(\overline{S}  D_1- D_1 \overline{R}\;\Gamma_5 D_1 + D_1 S - D_1\Gamma_5 R D_1\right)\psi_1. 
\end{gather}
The action will be invariant under this transformation if
\begin{gather}
\overline{S}  D_1- D_1 \overline{R}\;\Gamma_5 D_1 + D_1 S - D_1\Gamma_5 R D_1 = 0.
\end{gather}
This is the Ginsparg-Wilson relation, equation (\ref{eq:GW}), with
\begin{align}
\overline{S} =& \overline{B}\; \Gamma_5 \hat{\overline{B}},\nonumber\\
S = &\hat{B}\Gamma_5 B,\nonumber\\
\overline{R} = & \alpha^{-1} \overline{B}\; \Gamma_5 \hat{\overline{B}}\Gamma_5,\nonumber\\
R = & \Gamma_5 \hat{B} \Gamma_5 B \alpha^{-1}.\label{eq:19}
\end{align}
Thus this fermion action, derived from the continuum action by the block transformations outlined in the previous section, satisfies a Ginsparg-Wilson chiral symmetry. This is a generalisation of L\"uscher's original lattice chiral symmetry, which assumed that $[B,\gamma_5]=0$ and consequently had $S = \overline{S} = \gamma_5$. By considering the U(1) anomaly, I can easily derive a topological charge associated with this chiral symmetry,
\begin{gather}
Q_f = \frac{1}{2}\Tr\;\left[\overline{S} +  S - D_1 \overline{R}\gamma_5 - \gamma_5 R D_1\right].
\end{gather}
With $\alpha = \infty$ and therefore $R=0$ the topological charge is
\begin{gather}
Q_f = \frac{1}{2}\Tr\;\left[\overline{S} +  S\right] = \frac{1}{2}\Tr\;[ \hat{B}(\Gamma_5) B +  \overline{B}\; \Gamma_5 \hat{\overline{B}}].
\end{gather}
From the cyclicity of the trace, assuming that $\overline{B}$ and $B$ exist and are invertible, we have $Q_f = \Tr\;(\Gamma_5 + \gamma_5|\psi_0\rangle\langle\psi_0|)$\footnote{As noted in an earlier footnote, the additional $|\psi_0\rangle\langle\psi_0|$ is inserted using freedom provided by the zero mode symmetry of the Euclidean Lagrangian to allow continuation to Minkowski space.}. It is easy to show that (irrespective of whether we take the limit $\eta\rightarrow 0$)\footnote{Again, expressing $\gamma_5$ in the eigenvector basis of $D_0$.}
\begin{gather}
Q_f=\Tr\;\left[\gamma_5 - \frac{i\eta}{D_0 + i\eta\gamma_5} + \gamma_5|\psi_0\rangle\langle\psi_0|\right]=\text{index}(D_0).\nonumber
\end{gather}
This is the well-known Atiyah-Singer theorem~\cite{Atiyah-Singer} for QCD.

\section{Wilson fermions}\label{sec:4}
\begin{figure}[t]
\begin{center}
\begin{tabular}{c}
\psfrag{r}[lc][lc][0.75][0]{$(\overline{\psi}_1\psi_1)/(\overline{\psi}_0\psi_0)$}
\psfrag{xlabel}[bc][bc][0.75][0]{$x/a$}
\includegraphics{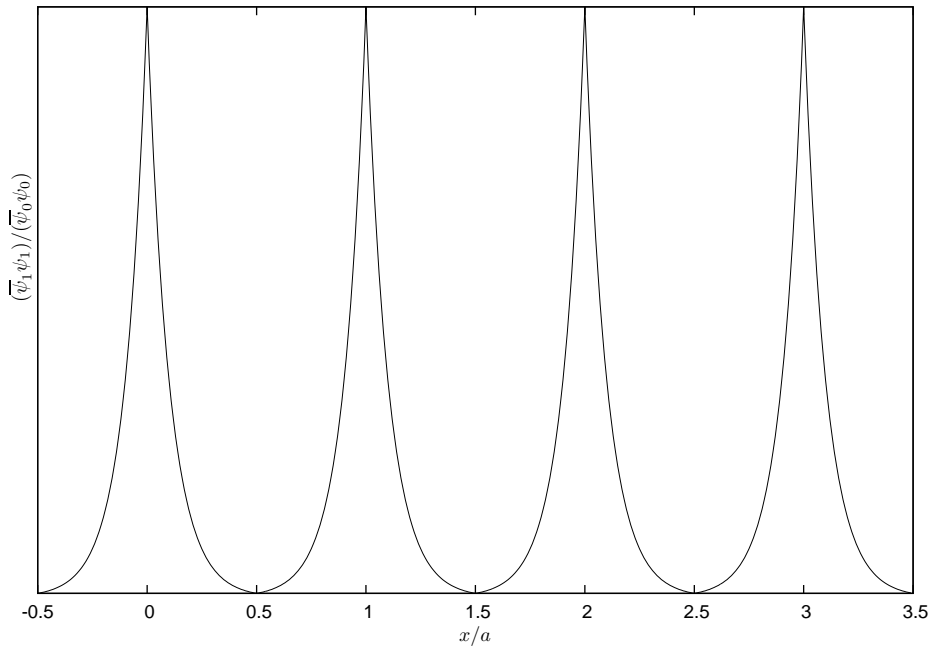}
\end{tabular}
\end{center}
\caption{An illustration of the method used in this approach. The ratio of the blocked fermion field, $\psi_1$, to the original field, $\psi_0$ is plotted as a function of the $x$-position in terms of lattice sites.}\label{fig:rough_plot}
\end{figure} 
\subsection{Introduction}
My intention is to block from a continuum fermion field to another continuum fermion field which will reduce to the lattice theory in a particular limit.  A crude one-dimensional example of what I am trying to achieve is illustrated in figure \ref{fig:rough_plot}.

The blocking will be a function of the $\gamma$-matrices and the gauge field $A_{\mu}$, but will also depend on two parameters, the lattice spacing $a$ and a second parameter $\zeta^{-1}$ which controls the width of the peaks in the blocked fermion field around the lattice sites.  As the width decreases to zero, which is controlled by the limit that $\zeta\rightarrow \infty$, I recover a  lattice action because only the fermion fields on the lattice sites contribute to the action. The blockings are constructed so that the integral over the various peaks in the fermion field remains finite even at small width, so that when $\zeta = \infty$ the new fermion field can be described using a sum over Dirac $\delta$-functions, and a lattice action will be recovered. The integral over space time in the action will thus become a sum over lattice sites. Clearly, until this limit is taken, the new Dirac operator remains invertible and well defined in the continuum; in particular the number of degrees of freedom for the blocked field are the same as for the original field. This is where this approach differs from previous renormalisation group blockings, which generally change the number of degrees of freedom by using rectangular blocking matrices. But once the limit is taken, at the end of the calculation, we will have a lattice theory. The only difficulty is in finding a blocking which, firstly, generates a particular lattice fermion action and, secondly, remains valid in the lattice limit according to the rules laid down in section \ref{sec:2}. In this section, I discuss a blocking which will generate the  Wilson fermion action, and in the subsequent sections a blocking which will generate the overlap action.     

\subsection{The blocking}
Consider the blocking
\begin{align}
\alpha(x,y) =& \Lambda\delta(x-y),\nonumber\\
B_W(y,x) = &\sum_n \zeta^4 e^{-\zeta \sum_{\mu}|x_{\mu} -a n_{\mu}|}
\prod_{\beta,\gamma}\theta\left(\frac{1}{2}a - |x_{\gamma} - an_{\gamma}|\right)
\theta\left(\frac{1}{2}a - |y_{\beta} - an_{\beta}|\right)
\nonumber\\
&\left(1 + r \sum_{\theta} \gamma_{\theta}N(y_{\theta} -a n_{\theta})\right)\sum_{\mathfrak{L}_{y,an},\mathfrak{L}_{an,x}}e^{-W[\mathfrak{L}_{y,an}]}U[\mathfrak{L}_{y,an}]e^{-W[\mathfrak{L}_{an,x}]}U[\mathfrak{L}_{an,x}],\nonumber 
\\
\overline{B}_W(x,y) = &\sum_n \zeta^4 e^{-\zeta \sum_{\mu}|x_{\mu} -a n_{\mu}|}
\prod_{\beta,\gamma}\theta\left(\frac{1}{2}a - |x_{\gamma} -a n_{\gamma}|\right)\theta\left(\frac{1}{2}a - |y_{\beta} -a n_{\beta}|\right)\nonumber\\
&\left(1 - r \sum_{\theta} \gamma_{\theta}N(y_{\theta} - an_{\theta})\right)\sum_{\mathfrak{L}_{x,an},\mathfrak{L}_{an,y}}e^{-W[\mathfrak{L}_{x,an}]}U[\mathfrak{L}_{x,an}]e^{-W[\mathfrak{L}_{an,y}]}[\mathfrak{L}_{an,y}],\label{eq:26}
\end{align}
where $W$ and $U$ are defined below; $\mathfrak{L}_{x,an}$ represents a path between continuum positions $x$ and $an$ and the sum is over all possible continuous paths\footnote{The precise definition of ``all possible continuous paths" is unimportant as long as, for each path contained within the sum, the path ordered integration over the gauge fields defined in equation (\ref{eq:30U}) remains differentiable, starts and terminates at the required locations, and contains the shortest path between the two points. It could, for example, either be defined to be the paths bound within the hypercube around $x$ or not. It would also be possible to remove this sum and just consider the direct path; however the construction used here is more general and will also permit an easy modification to, for example, allow smeared links.}; $\zeta$ is a tunable parameter, where I define the `lattice limit' as $\zeta\rightarrow\infty$ (if such a limit exists); and $r>1$ is another tunable parameter, which will be related to the coefficient of the Wilson term in the final action. $n_{\mu}$ is restricted to integer values. Throughout this work, I shall frequently suppress the lattice spacing by setting $a=1$. I shall always use $\Lambda=\infty$. Here I have defined the functions
\begin{gather}
\theta(x) =\left\{\begin{array}{l l}
0 &x<0\\
1 &x > 0\\
\frac{1}{2}&x=0\end{array}
\right. ,
\end{gather}
and
\begin{gather}
N(x) =\left\{\begin{array}{l l}
0 &|x|<(1-\epsilon)\frac{a}{2}\\
\text{sign}(x) &|x| > (1-\epsilon)\frac{a}{2}\\
\frac{1}{2}\text{sign}(x)&|x|=(1-\epsilon)\frac{a}{2}
\end{array}
\right. ,
\end{gather}
where $\epsilon$ is some tunable parameter in the range $0<\epsilon\le1$.  I will also define
\begin{align}
W[\mathfrak{L}_{x,n}] =& \zeta \left[2\int_{\mathfrak{L}} {s}_{\mu} d{s}_{\mu} - \sum_{\mu}\prod_{\nu\neq\mu}\theta\left((x_{\mu}-n_{\mu})^2 - (x_{\nu}-n_{\nu})^2\right)(x_{\mu} - n_{\mu})^2\right]^{\frac{1}{2}} + \log \xi,\\
\intertext{and}
U[\mathfrak{L}_{x,n}] = & P\left[e^{-ig\int_{\mathfrak{L}}A_{\mu}({s})d{s}_{\mu}}\right],\label{eq:30U}
\end{align} 
where $P$ represents path-ordering, $U[\mathfrak{L}_{x,n}] = U^{\dagger}[\mathfrak{L}_{n,x}]$, ${s}_{\mu}$ represents a position in space-time, and the normalisation constant $\xi$ is chosen so that 
\begin{gather}
\sum_{\mathfrak{L}}e^{-W[\mathfrak{L}]} = e^{-\zeta a/2 (\sum_{\mu}|x_{\mu} - n_{\mu}|)/\left(\sum_{\mu}\prod_{\nu\neq\mu}(x_{\mu} - n_{\mu})\theta((x_{\mu} - n_{\mu})^2 -(x_{\nu} - n_{\nu})^2 
)\right)}.\nonumber
\end{gather}
The precise form of $W$ is unimportant, as long as it is an even function of $x-n$ and the direct path between $x$ and $n$ and the path along the axes of the lattice dominate at large $\zeta$. The presence of the path ordered gauge links within $U$ ensures that this blocking, and any constructed from it (such as those in sections \ref{sec:5} onwards), is gauge-covariant, satisfying the fourth condition required for a valid blocking. Note that $N$ is anti-hermitian and thus $\overline{B}(x,y) = (B(y,x))^{\dagger} = \gamma_5 B(x,y)\gamma_5$, where the Hermitian conjugate acts on the spatial indices.

In equation (\ref{eq:26}), the continuum Dirac operator $D_0$ will act on the position $y$ while $x$ relates to the space time coordinate of the spinor field $\psi_1$. The $\zeta^4 e^{-\zeta |x - n|}$ term insures that $\psi_1$ is dominated by the lattice contributions. The $\theta$ terms restrict the lattice fields to  the hypercube centred on the lattice site, while $1-r\gamma_{\theta} N(y_{\theta}- n_{\theta})$ will generate the Wilson term in the final action. The terms depending on the gauge fields ensure that gauge covariance is satisfied. 

The inverse blockings are
\begin{align}
B_W^{-1}(x,y) = &\sum_n \mathfrak{N}_n(x,x')\zeta^{-4} e^{\zeta \sum_{\mu}|x'_{\mu} - n_{\mu}|}\prod_{\beta,\gamma}\theta\left(\frac{1}{2} - |x'_{\gamma} - n_{\gamma}|\right)\theta\left(\frac{1}{2} - |y'_{\beta} - n_{\beta}|\right)\nonumber\\
&\left(1 - r \sum_{\theta}\gamma_{\theta} N(y_{\theta} - n_{\theta})\right)\nonumber\\&\sum_{\mathfrak{L}_{n,y'},\mathfrak{L}_{x',n}}e^{-W[\mathfrak{L}_{x',n}]}U[\mathfrak{L}_{x',n}]e^{-W[\mathfrak{L}_{n,y'}]}U[\mathfrak{L}_{n,y'}] \mathfrak{N}_n(y',y),\nonumber
\\
\overline{B}_W^{-1}(y,x) = &\sum_n \mathfrak{N}_n(y,y')\zeta^{-4} e^{\zeta \sum_{\mu}|x_{\mu} - n_{\mu}|}\prod_{\beta,\gamma}\theta\left(\frac{1}{2} - |x_{\gamma} - n_{\gamma}|\right)\theta\left(\frac{1}{2} - |y_{\beta} - n_{\beta}|\right)\nonumber\\
&\left(1 + r \sum_{\theta} \gamma_{\theta}N(y_{\theta} - n_{\theta})\right)\nonumber\\&\sum_{\mathfrak{L}_{n,x'},\mathfrak{L}_{n,y}}e^{-W[\mathfrak{L}_{y',n}]}U[\mathfrak{L}_{y',n}]e^{-W[\mathfrak{L}_{n,x'}]}U[\mathfrak{L}_{n,x'}]\mathfrak{N}_n(x',x).\label{eq:inverseBlockings}
\end{align}
$\mathfrak{N}$ is a normalisation constant constructed from gauge fields, 
\begin{gather}
\mathfrak{N}^{-1}_n(x',x) =  \sum_{\mathfrak{L}_{x',n},\mathfrak{L}_{n,x}}e^{-W[\mathfrak{L}_{x',n}]} U[\mathfrak{L}_{x',n}] e^{-W[\mathfrak{L}_{n,x}]} U^{\dagger}[\mathfrak{L}_{n,x}].
\end{gather}
In the lattice limit, $\mathfrak{N}_n^{-1}(x,x') = 1$.

From this blocking, we can construct a new Dirac operator $D_1$. First of all, I define $D_0$ as the continuum operator
\begin{gather}
D_0 F(y) = \sum_{\mu}\gamma_{\mu}e^{ig\int^y A_{\nu}({s})d{s}_{\nu}}\partial_{\mu}\left(e^{-ig\int^y A_{\nu}({s})d{s}_{\nu}}F(y)\right).
\end{gather}
Then, 
\begin{align}
D_0B_W(y,x) =& -\gamma_{\mu}\zeta^4\prod_{\beta}\sum_{\mathfrak{L},n}e^{-W[\mathfrak{L}_{y,n}]}U[\mathfrak{L}_{y,n}]e^{-W[\mathfrak{L}_{n,x}]}U[\mathfrak{L}_{n,x}] e^{-\zeta|x_{\mu} - n_{\mu}|}\nonumber\\
&\Bigg[\sum_{\nu\neq\mu} \theta\left(\frac{1}{2} - |y_{\nu} - n_{\nu}|\right)\delta\left(\frac{1}{2} - |y_{\mu} - n_{\mu}|\right)\text{sign}(y_{\mu} - n_{\mu})(1+r\gamma_{\theta}N_{\theta})-\phantom{a}\nonumber\\& 
 \prod_{\nu}\theta\left(\frac{1}{2} - |y_{\beta} - n_{\beta}|\right)r\gamma_{\mu}\delta\left(|y_{\mu} - n_{\mu}| - \frac{1-\epsilon}{2}\right)+\nonumber\\
&\partial_{\mu}W[\mathfrak{L}_{y,n}](1+r\gamma_{\theta}N_{\theta})\theta\left(\frac{1}{2} - |y_{\beta} - n_{\beta}|\right)
\Bigg]\theta\left(\frac{1}{2} - |x_{\beta} - n_{\beta}|\right), 
\end{align}
where I write $N_{\theta}$ as a shorthand for $N(y_{\theta} - n_{\theta})$.
From this, I obtain
\begin{align}
\overline{\psi}_1D_1\psi_1=&\overline{\psi}_1(x')\overline{B}_W(x',y)D_0B_W(y,x)\psi_1(x) \nonumber\\
=&\zeta^8\sum_{n,n'}\overline{\psi}_1(x') e^{-\zeta |x'_{\nu'} - n'_{\nu'}|}\theta\left(\frac{1}{2} - |x'_{\gamma'} - n'_{\gamma'}|\right)e^{-\zeta |x_{\nu} - n_{\nu}|}
\nonumber\\
&\theta\left(\frac{1}{2} - |x_{\gamma} - n_{\gamma}|\right)\int d^4y\sum_{\mathfrak{L}} e^{-W[\mathfrak{L}_{x',n'}]}U[\mathfrak{L}_{x',n'}]e^{-W[\mathfrak{L}_{n',y}]}U[\mathfrak{L}_{n',y}]\nonumber\\
&\phantom{somespace}D_W^{n',n}(y)e^{-W[\mathfrak{L}_{y,n}]}U[\mathfrak{L}_{y,n}]e^{-W[\mathfrak{L}_{n,x}]}U[\mathfrak{L}_{n,x}] \psi_1(x),
\end{align}
where 
\begin{align}
&D_W^{n',n}(y) = \nonumber\\
&\frac{1}{2}\delta_{n,n'}(\gamma_{\mu} + 2 r \gamma_{\mu}\gamma_{\theta}N_{\theta} + \gamma_{\mu}r^2N_{\theta}^2 - 2r^2N_{\theta}\gamma_{\theta}N_{\mu} - r^2\gamma_{\mu}N_{\mu}^2) \theta\left(|y_{\nu} - n_{\nu}| - \frac{1}{2}\right)\nonumber\\
&\phantom{space}\left(\delta\left(y_{\mu} - n_{\mu} - \frac{1}{2}\right) - \delta\left(y_{\mu} - n_{\mu} + \frac{1}{2}\right)\right)+\nonumber\\
& \frac{1}{2}\delta_{n'+\mu,n}  (\gamma_{\mu}(1+r^2 + N_{\theta}^2r^2) - 2r + 2rN_{\theta}\gamma_{\mu}\gamma_{\theta})\delta\left(y_{\mu} - n_{\mu} + \frac{1}{2}\right)\theta\left(|y_{\nu} - n_{\nu}| - \frac{1}{2}\right) + \nonumber\\
&\frac{1}{2}\delta_{n'-\mu,n} (-\gamma_{\mu}(1+r^2 + N_{\theta}^2r^2) - 2r - 2rN_{\theta}\gamma_{\mu}\gamma_{\theta}) \delta\left(y_{\mu} - n_{\mu} - \frac{1}{2}\right)\theta\left(|y_{\nu} - n_{\nu}| - \frac{1}{2}\right) + \nonumber\\
&
\delta_{n,n'}r(1-r\gamma_{\theta}N_{\theta}-r\gamma_{\mu}N_{\mu})\theta\left(|y_{\nu} - n_{\nu}| - \frac{1}{2}\right)\left(\delta\left(y_{\mu} - n_{\mu} - \frac{1-\epsilon}{2}\right) + \delta\left(y_{\mu} - n_{\mu} + \frac{1-\epsilon}{2}\right)\right)+
\nonumber\\
&\delta_{n,n'}\partial_{\mu} W[\mathfrak{L}_{y,n}](\gamma_{\mu} + 2 r \gamma_{\mu}\gamma_{\theta}N_{\theta} + \gamma_{\mu}r^2N_{\theta}^2 - 2r^2N_{\theta}\gamma_{\theta}N_{\mu} - r^2\gamma_{\mu}N_{\mu}^2) \theta\left(|y_{\nu} - n_{\nu}| - \frac{1}{2}\right),
\end{align}
where sums over $\mu$, $\theta\neq \mu$ and $\nu\neq\mu$ are assumed. 
The first three terms are obtained from the differential of $\theta$, the fourth from the differential of $N$, and the last from the differential of $W$. $N$ and $\partial_{\mu}W$ are odd functions of $y$, so, if the gauge fields are sufficiently smooth, the contributions to $D_W$ from terms odd in $N$ or $\partial_{\mu}W$ will be suppressed by powers of the lattice spacing, and in the free theory they will not contribute. Neither will they contribute In the lattice limit: as $\zeta\rightarrow\infty$ only the shortest paths of gauge links will survive, and in particular 
\begin{gather}
\sum_{\mathfrak{L}_{ny}} e^{-W[\mathfrak{L}_{ny}]}U[\mathfrak{L}_{ny}]\sum_{\mathfrak{L}_{yn}}e^{-W[\mathfrak{L}_{yn}]}U[\mathfrak{L}_{yn}] =0\nonumber
\end{gather}
for all $y$ except along one of the axes of the lattice, where it will be 1. Since this is the only dependence on $y$ within the integral except within $D_W$, those terms in $D_W$ which are odd in any component of $y$ must cancel.
This means that the for those terms in $D_W$ proportional to $\delta_{n,n'}$ the only dependence on $y-n$ in the expression for $\overline{\psi}_1D_1\psi_1$ will come from $D_W$, and the integration over all the odd functions of $(y-n)$ in $D_W$ will give zero. I can define, 
\begin{align}
U_{\mu}(n) = \int d^4y \theta\left(\frac{1}{2} - | y_{\nu} - n_{\nu}|\right) &\delta\left(\frac{1}{2} - y_{\mu} + n_{\mu}\right)e^{-W[\mathfrak{L}_{n,y}]}e^{-W[\mathfrak{L}_{y,n}]}\nonumber\\
&P\left[e^{-ig\int_{\mathfrak{L}_{n,y}}A_{\nu}({s})d{s}_{\nu}}\right]
P\left[e^{ig\int_{\mathfrak{L}_{y,n+\hat{\mu}}}A_{\nu}({s})d{s}_{\nu}}\right].
\end{align}
Since as $\zeta \rightarrow \infty$, $e^{-W}$ only survives for direct paths where $y$ lies along one of the four Cartesian axes of the lattice, $U_{\mu}(n)$ becomes the path ordered gauge transporter along the direct path between $n$ and $n+\mu$, which is the standard definition of the link in lattice gauge theory.   
 Furthermore, the action will be dominated by the spinor fields at the lattice sites, so that we can write,   
\begin{align}
\overline{\psi}_1D_1\psi_1 =&
 \frac{1}{2}\sum_n\overline{\psi}_1(n+\hat{\mu})(-2r+(1+(1+3\epsilon)r^2)\gamma_{\mu})U^{\dagger}_\mu(n)\psi_1(n) +\nonumber\\& \overline{\psi}_1(n-\hat{\mu})(-2r-(1+(1+3\epsilon)r^2)\gamma_{\mu}) U_{\mu}(n-\hat{\mu})\psi_1(n) + (8r) \overline{\psi}(n)\psi(n),
\end{align}
which, up to some normalisation factor and redefinition of terms, is the standard lattice Wilson action.

We can take the Fourier transform of this continuum Wilson operator, which, in the free theory, gives
\begin{align}
\FT{D_1}(p) =& \frac{1}{\int d^4x}\int d^4x d^4x' e^{-ipx'} D_1(x',x) e^{ipx}\nonumber\\
=&\prod_{\mu}\left(\frac{2\zeta^2}{\zeta^2 + p_{\mu}^2} + e^{-\zeta/2}\frac{2\zeta p_{\mu}\sin(p_{\mu}/2) + 2 \zeta^2 \cos(p_{\mu}/2)}{\zeta^2 + p_{\mu}^2}\right)^2 \nonumber\\&\sum_{\nu} (i \gamma_{\nu}(1+r^2 + 3\epsilon r^2)(\sin(p_{\nu})) +2r(1-\cos(p_{\nu}))). 
\end{align}
As $\zeta\rightarrow\infty$ this becomes, again up to a normalisation factor and redefinition of variables, the familiar expression for Wilson fermions.

Of course, this blocking transformation, although it generates the Wilson fermion action, is not valid in the lattice limit. This can be seen by considering the Fourier transform of $\overline{B}_W^{\phantom{.}-1}B_W^{-1}$, which, in the free field approximation, gives
\begin{align}
&\FT{(B_W\overline{B}_W)^{-1}}(p) =\nonumber\\&\phantom{spa} \prod_{\mu}\left(-\frac{2}{\zeta^2 + p_{\mu}^2} + e^{\zeta/2}\frac{2 p_{\mu}\sin(p_{\mu}/2)/\zeta + 2 \cos(p_{\mu}/2)}{\zeta^2 + p_{\mu}^2}\int d^4y(1-r^2N_{\theta}^2)\right)^2 .
\end{align} 
This is not analytic in the limit that $\zeta\rightarrow\infty$, hence, using the Paley-Wiener theorem~\cite{Paley-Wiener},  $\overline{B}_W^{\phantom{.}-1}B_W^{-1}$ is not local (at least in the free case, which strongly suggests that it will also not be local in the interacting theory), which means that this blocking fails the conditions outlined in section \ref{sec:2}. It seems intuitively obvious that a similar picture will  hold for any operator which restricts the spinor fields to the lattice sites: the inverse blocking for any lattice Dirac operator which projects off-lattice site elements to zero must be infinite for any position that is not on a lattice site, and from this one would suppose that the Fourier transform of the inverse operator would be non-analytic. However, this is not always the case. 

\section{Overlap fermions}\label{sec:5}
In this and the following sections I shall use, in addition to the standard $\gamma$-matrix representation, an additional, somewhat perverse, representation of the $\gamma$-matrices, which I shall label $g_{\mu}$: a function of the Dirac operators. The relationship between these two $\gamma$-matrix representations is given in appendix \ref{app:B}, together with an outline of why I need to use it. Here I just note that $g_5 = \gamma_5$ is diagonal, that $g_{\mu}$ satisfies the same anti-commutation relationships as $\gamma_{\mu}$, and transforms in the same way under $\mathcal{CP}$, and relegate all other details to the appendix.
\subsection{Introduction}
In this section, I will use a modified form of the Wilson blocking of the previous section:
\begin{align}
B_W(y,x) = &\sum_n \zeta^4 e^{-\zeta \sum_{\mu}|x_{\mu} - n_{\mu}|}
\prod_{\beta,\gamma}\theta\left(\frac{1}{2} - |x_{\gamma} - n_{\gamma}|\right)
\theta\left(\frac{1}{2} - |y_{\beta} - n_{\beta}|\right)
\nonumber\\
&e^{-m\gamma_{\mu}(1+r^2 + 3\epsilon r^2)(y_{\mu}-n_{\mu})/(1+2r^2\epsilon)}\left(1 + r \sum_{\theta} \gamma_{\theta}N(y_{\theta} - n_{\theta})\right)\nonumber\\ &
\sum_{\mathfrak{L}_{x,n},\mathfrak{L}_{n,y}}e^{-W[\mathfrak{L}_{x,n}]}U[\mathfrak{L}_{xn}]e^{-W[\mathfrak{L}_{n,y}]}U[\mathfrak{L}_{n,y}] ,\nonumber
\\
\overline{B}_W(x',y) = &\sum_n \zeta^4 e^{-\zeta \sum_{\mu}|x_{\mu} - n_{\mu}|}
\prod_{\beta,\gamma}\theta\left(\frac{1}{2} - |x_{\gamma} - n_{\gamma}|\right)\theta\left(\frac{1}{2} - |y_{\beta} - n_{\beta}|\right)\nonumber\\
&\left(1 - r \sum_{\theta} \gamma_{\theta}N(y_{\theta} - n_{\theta})\right)e^{m\gamma_{\mu}(1+r^2 + 3\epsilon r^2)(y_{\mu}-n_{\mu})/(1+2r^2\epsilon)}\nonumber\\
&\sum_{\mathfrak{L}_{x,n},\mathfrak{L}_{n,y}}e^{-W[\mathfrak{L}_{x,n}]}U[\mathfrak{L}_{xn}]e^{-W[\mathfrak{L}_{n,y}]}U[\mathfrak{L}_{n,y}].\label{eq:45}
\end{align}
The additional term $e^{-m\gamma_{\mu}(1+r^2 + 3\epsilon r^2)(y_{\mu}-n_{\mu})/(1+2r^2\epsilon)}$ has the effect of introducing a negative mass of magnitude $m$; the value of this mass is constrained between the critical Wilson mass and 2  to ensure that there are no doublers in the overlap action. To avoid complications arising from the non-analyticity of the inverse of the Dirac operator, I use the $i\eta\gamma_5$ mass regularisation discussed in section \ref{sec:3.1}, for both Dirac operators, the original $D_0$ and the new $D_2$. Additionally, I shall introduce the blocking operator
\begin{align}
B_C(y,x) = \sum_n& \theta\left(\frac{1}{2} - |y_{\beta} - n_{\beta}|\right)\theta\left(\frac{1}{2} - |x_{\beta} - n_{\beta}|\right) \sum_{\mathfrak{L}} e^{-W[\mathfrak{L}_{yn}]}U[\mathfrak{L}_{yn}]\nonumber\\
&e^{-W[\mathfrak{L}_{nx}]}U[\mathfrak{L}_{nx}](\delta^4(y-x) - \zeta^4e^{-\zeta|y-n|}e^{-\zeta|x-n|}) .
\end{align}
It is easy to demonstrate that $D_1 B_C = O(e^{-\beta\zeta})$ and that 
\begin{gather}
B_C^{m} = B_C \left(\int d^4 x \theta\left(\frac{1}{2} - |x_{\beta} - n_{\beta}|\right) e^{-W[\mathfrak{L}_{nx}]}U[\mathfrak{L}_{nx}]e^{-W[\mathfrak{L}_{xn}]}U[\mathfrak{L}_{xn}]\right)^m + O(e^{-\beta'\zeta}),
\end{gather}
where $\beta$ and $\beta'$ are positive and real. $B_C$ will be used to ensure that the final action is dominated by the fermion fields on the lattice sites.
I will use a blocking constructed from $\gamma_5$, $B_W$,$\overline{B}_W$, $B_C$ and $D_0$ as follows:
\begin{align}
\overline{B} =& Z^{\dagger}\nonumber\\
B =&\zeta^4\frac{1}{D_0 + i\eta\gamma_5}Z( 1+i\mathbb{I} \eta\gamma_5 + \gamma_5 F(\gamma_5 (D_1 - B_C)))\nonumber\\
D_1=&\overline{B}_W D_0 B_W,\label{eq:overlapblocking1}
\end{align}
where $F$ is an arbitrary real function and $\mathbb{I}$ is the lattice identity operator given by.
\begin{align}
\mathbb{I} = \sum_n\zeta^4\theta\left(\frac{1}{2} - |x-n|\right)&\theta\left(\frac{1}{2} - |y-n|\right)\nonumber\\
& \sum_{\mathfrak{L}} e^{-W[\mathfrak{L}_{xn}]}U[\mathfrak{L}_{xn}]
e^{-W[\mathfrak{L}_{ny}]}U[\mathfrak{L}_{ny}]e^{-\zeta|x_{\mu} - n_{\mu}|}e^{-\zeta|y_{\mu} - n_{\mu}|}.
\end{align}
This blocking will generate a Dirac operator $D_2 = 1+\gamma_5 F(\gamma_5 (D_1-B_C))+i\mathbb{I}\eta\gamma_5$. $Z$ is a unitary operator, and initially I shall work in the trivial topological sector where it is possible to set $Z=1$. In non-trivial topological sectors, it is necessary to use a different form of $Z$ for reasons which shall be discussed later.  To ensure that $D_2$ transforms correctly under $\mathcal{CP}$, each term within its expansion in the lattice spacing must contain an even number of $\gamma_5$s, which means that $F$ must be an odd function of $\gamma_5 (D_1-B_C)$ (see appendix \ref{app:CP1b}).

Although I shall proceed as far as possible using a general $F$, my ultimate aim is to demonstrate that the blocking is valid for the overlap operator, where $F(x) = \sign(x)$. Therefore I shall consider this case when it is necessary to move from the general argument to a specific example. 

To demonstrate that this is a valid blocking, I need to show that
\begin{enumerate}
\item $\hat{B}$ and $B$ exist, i.e. are not zero or infinite;
\item $B$ and $\hat{B}$ are local. 
\item $\hat{B}\hat{\overline{B}}$ has a positive real part;
\end{enumerate}
\subsection{Existence of $\hat{B}$ and $B$}
For the blocking and inverse blocking to exist, two conditions must be satisfied: firstly, $F(x)$ must remain finite for all $x$ (which has to be tested for specific examples), and secondly the blocking and inverse blockings must not have any zero modes. This second condition follows from the positivity of the blockings, which is discussed in section \ref{sec:positivity}. As discussed in section \ref{sec:2}, this only holds if $\text{Index}(D_0) = \text{Index}(D_2)$ and $D_2$ has no doublers, which includes overlap fermions under certain conditions.
\subsection{Locality of $B$}\label{sec:5.3}
To demonstrate the locality of the blocking, I calculate the Fourier transform. By the Paley-Wiener theorem~\cite{Paley-Wiener}, if the Fourier transform is analytic along the real axis, then the blocking is local. I proceed by expanding  the function $F$ in terms of a polynomial of the Hermitian Wilson operator, which will be valid (in the case of overlap fermions) as long as $\gamma_5( D_1 - B_C)$ has no eigenvalues which are exactly zero.
\begin{align}
B = \zeta^4
\frac{1}{D_0+i\gamma_5 \eta}\bigg( 1+ i\mathbb{I}\gamma_5\eta +& \gamma_5 \sum_m c_m (\gamma_5\overline{B}_W D_0 B_W)^m +
\gamma_5 \sum_m c_m (-\gamma_5 B_C)^m + O(e^{-\alpha\zeta})\bigg).
\end{align}
The Fourier transform, $\FT{B_T}$, of the product of Wilson operators is
\begin{align}
\FT{B_T} = \prod_{i=0}^m\left( \int d^4 x_i\right)\prod_{i = 0}^{m-1}&\left(d^4y_i \sum_{m_i}\right)e^{-ipx_0}
\prod_{i = 0}^{n-1}\left({\overline{B}_W}_{x_i,y_i,n_i} {D_0} {B_W}_{y_i,n_i,x_{i+1}} \right) e^{ipx_n}. 
\end{align}
$B_W$ has been designed so that it can be factorised into $B_W(x,y) = \sum_n B_y(n,y)B_x(x,n)$, and similarly for $\overline{B}_W$. Then just three integrals are needed to calculate the Fourier transform:
\begin{align}
e_n(p) =& \int d^4 x e^{ip(x-n)}B_x(x,n)\nonumber\\
x_n =&\int d^4 x B_x(x,n)\overline{B}_x(n,x)\nonumber\\
d_n(p) = &\int d^4y e^{-ipn'}\overline{B}_y(n',y)D_0 B_y(y,n)  e^{ipn}.\label{eq:Ireallydontwanttocalculatethis}
\end{align}
In the free theory, these functions are given by
\begin{align}
e_n = & \prod_{\mu}\left(\frac{2\zeta^2}{p_{\mu}^2 + \zeta^2} + e^{-\zeta/2}\frac{2 p_{\mu} \zeta (\sin (p_{\mu}/2)) - 2 \zeta^2 \cos (p_{\mu}/2)}{\zeta^2 + p^2_{\mu}}\right)
\\
x_n = &(1-e^{-\zeta})^4\zeta^4
\\
d_n = & \sum_{\mu}\left[i\gamma_{\mu} \sin (p_{\mu})(1+r^2 + 3\epsilon r^2) + 2r(1-\cos(p_{\mu}/2) )\right] - m(1+r^2 +3\epsilon r^2).
\label{eq:free theory}
\end{align}
The Fourier transform of the polynomial series in the Wilson operator is
\begin{gather}
\FT{B_T} = \sum_n (d_n x_n)^m \frac{e_n e_n^{\dagger}}{x_n}.
\end{gather}
The Fourier transform of the lattice identity operator 
is $\prod_{\mu}(2 \sin (p_{\mu}/2)/p_{\mu})$, and the momentum representation of $B_C$ is 
\begin{gather}
\FT{B_C} = \sum_n\left[\prod_{\mu}\left(\frac{2 \sin (p_{\mu}/2)}{p_{\mu}}\right) - \frac{1}{\zeta^4}e_n e_n^{\dagger} \right]
\end{gather} 
If $d_n x_n$ remains within the radiance of convergence for the polynomial, the momentum representation of the blocking will be
\begin{gather}
\FT{B} = \FT{\frac{1}{D_0+i\gamma_5 \eta}}\left(e_n e_n^{\dagger}(1+i\gamma_5 \eta) + \zeta^4 \sum_n  \frac{e_n e_n^{\dagger}}{x_n}\gamma_5 F(\gamma_5 x_n d_n)\right) + O(e^{-\alpha\zeta}).
\end{gather}
At small $p$, $1+\gamma_5 F(\gamma_5x_n d_n)$  can be expanded
\begin{gather}
1+\gamma_5 F(\gamma_5x_n d_n) = 1 + \gamma_5c_0 + x_n(p)d_n(p)c_1 + c_2 x_n(p) d_n(p) \gamma_5 x_n d_n(p) + \ldots, 
\end{gather}
where $x_n(p) d_n(p) = \FT{D_0}(p) - m + O(\FT{D_0}(p)^2)$. If $\FT{B}$ is to be analytic, the series expansion of $1+\gamma_5 F$ must become $\FT{D_0}(p) + O(\FT{D_0}(p)^2)$ close to the zeros of $\FT{D_0}(p)$.

If $F$ is chosen so that at small $p$, $\FT{D_2}(p) =\FT{D_0}(p) + O(p^2)$, which in the free theory corresponds to $F(\gamma_5 x_n d_n)\sim \gamma_5(-1 + \gamma_{\mu}p_{\mu} + O(p^2))$, then $\FT{B}$ will remain analytic and so $B$ will be local. In the language of lattice gauge theory, this is equivalent to saying that the Dirac operator $D_2$ must have the correct continuum limit, which, as is well-known, overlap fermions do.
\subsection{Locality of $\hat{B}$}
The inverse blocking is defined as\footnote{So that I can easily perform the integrals, I define the inverse function $1/(1+\gamma_5F(\gamma_5 D_1))$ in terms of a polynomial expansion in $D_1$, but not $D_1^{-1}$, since the latter is not required for the purposes of this work. $\hat{B}(x',y)$ is defined so that $\int d^4y \hat{B}(x',y) B(y,x) = \zeta^8\sum_n e^{-\zeta|x-n|}e^{-\zeta|x'-n|}$; the limit of this function as $\zeta\rightarrow\infty$ is $\delta(x-x')$. It should be observed that this formulation of $\hat{B}$, because of the limitations of the polynomial series, is only strictly the inverse in the lattice limit; outside the lattice limit this `$\hat{B}$' will be the inverse of $B$ plus an additional term  which cannot contribute as $\zeta\rightarrow\infty$ and thus cannot cause any non-analyticity in the Fourier transform. When expanding $\hat{B}$ and $B$ in polynomial series, each integral over $B_x\overline{B}_x$ will give a factor of $\zeta^{-4}$, giving a total of $\zeta^{4-4m-4m'}$, where $m$ and $m'$ are the powers of the terms in the expansion under consideration. There are $2(m + m')$ factors of $\zeta^4$ in the definition of $D_1$, of which half are absorbed into the function $F$ (for example, when considering the Fourier transform, $x_n$ was proportional to $\zeta^4$). With a factor of $\zeta^8$ required for the $\delta$-function (so that the integral over the $\delta$-function is one), by counting powers of $\zeta$, I am left needing to insert one factor of $\zeta^4$ into either $B$ or $\hat{B}$, and I chose to insert it into $B$. Thus the definition of $\hat{B}$ as defined in equation (\ref{eq:59}) does not need to be multiplied by $\zeta^{-4}$.}
\begin{gather}
\hat{B} = (1+i\mathbb{I}\gamma_5 \eta + \gamma_5 F(\gamma_5 D_1 - \gamma_5 B_C))^{-1}Z^{\dagger} (D_0+i\eta\gamma_5).\label{eq:59}
\end{gather}

Once again, the Fourier transform can be calculated by expanding this in a polynomial series in $\gamma_5D_1 - \gamma_5 B_C$. It is necessary to treat the two cases $|i\gamma_5  \eta + \gamma_5 F|>1$ and $|i\gamma_5 \eta + \gamma_5 F|<1$ separately. For the second case, I expand in a geometric series:
\begin{align}
\hat{B}&(D_0+i\eta\gamma_5)^{-1} =\nonumber\\
& (1-i\mathbb{I}\eta\gamma_5 -\gamma_5F(\gamma_5 D_1 - \gamma_5 B_C) + (\gamma_5 F(\gamma_5 D_1 - \gamma_5 B_C)+  i\mathbb{I}\eta\gamma_5)^2 - \ldots).
\end{align}
Using the same technique as in the previous section, the Fourier transform of the blocking is
\begin{align}
\FT{\hat{B}} =\bigg(&\frac{2\sin(p_{\mu}/2)}{p_{\mu}} - \frac{e_n^{\dagger}e_n}{x_n}( ix_n\eta\gamma_5 + \gamma_5F(\gamma_5 x_n d_n))(1-(ixF_n\eta\gamma_5 + \gamma_5 F(\gamma_5 x_n d_n))\nonumber\\ &+ (ix_n\eta\gamma_5 + \gamma_5F(\gamma_5x_n d_n))^2 - \ldots\bigg) +\phantom{a}\nonumber\\
& \left(\frac{2\sin (p_{\mu}/2)}{p_{\mu}} - \frac{e_n e_n^{\dagger}}{\zeta^4}\right)(1+A+A^2+A^3+\ldots)\bigg)(D_0 + i\eta\gamma_5) + O(e^{-\zeta}),
\end{align}
where
\begin{align}
A =& \sum_n\int d^4x \theta\left(\frac{1}{2} - |x-n|\right)\theta\left(\frac{1}{2} - |x'-n|\right) e^{-W[\mathfrak{L}_{xn}]}U[\mathfrak{L}_{xn}]e^{-W[\mathfrak{L}_{nx'}]}U[\mathfrak{L}_{nx'}]\nonumber\\ =& 1-O(e^{-\alpha\zeta}).
\end{align}
Assuming that the geometric series converges (whether this is true depends on the precise form of $F$, however,  $|A|<1$ so the geometric series in $A$ will always converge), this gives,
\begin{align}
\FT{\hat{B}} = \bigg( - \frac{e_n^{\dagger}e_n}{x_n}&(ix_n\eta\gamma_5 + \gamma_5F(\gamma_5x_n d_n))\frac{1}{1+ix_n\eta\gamma_5 + \gamma_5 F(\gamma_5x_n d_n)} + \nonumber\\
&\prod_{\mu}\left(\frac{2\sin(p_{\mu}/2)}{p_{\mu}}\right)+\left[\prod_{\mu}\left(\frac{2\sin (p_{\mu}/2)}{p_{\mu}}\right) - \frac{e_n e_n^{\dagger}}{\zeta^4}\right](1-A)^{-1}\bigg)\nonumber\\
&\phantom{space}(\FT{D_0}+i\eta\gamma_5) + O(e^{-\zeta}),
\end{align}
and it is clear that in the limit $\zeta\rightarrow\infty$, $A\rightarrow 1$, $\FT{\hat{B}}\rightarrow\infty$ and thus $\hat{B}$ is not local.

However, if $|F + i\eta\gamma_5|>1$, one must use an alternative series expansion. By writing $G(D_1) = (i\mathbb{I}\eta\gamma_5 + \gamma_5 F(\gamma_5 D_1))^{-1}$, I express the inverse blocking as
\begin{align}
{B}^{-1} =&
(1-B_C + i\mathbb{I}\eta\gamma_5 + \gamma_5F(\gamma_5 D_1) + O(e^{-\zeta}))^{-1}(D_0 + i \eta\gamma_5)\nonumber\\
=&
 G((1-B_C)G + 1)^{-1}(D_0+i\eta\gamma_5) +O(e^{-\zeta}),\nonumber\\
\FT{\hat{B}}=&e_n e_n^{\dagger}G(x_n d_n)\frac{1}{G( x_n d_n) + 1}(\FT{D_0}+i\eta\gamma_5) + O(e^{-\zeta}),\nonumber\\
=&e_n e_n^{\dagger}(1+i\eta\gamma_5 + \gamma_5F(\gamma_5x_n d_n))^{-1}(\FT{D_0}+i\eta\gamma_5) + O(e^{-\zeta}).
\end{align}
I have assumed that $G$, like $F$, can be constructed from $D_1$ and the $\gamma$-matrices. The argument can easily be extended if $G$ is a function of $D_1$ and $D_1^{-1}$, but only if the Fourier transform of $(1-B_C)G$ is less than one despite the infinities coming from the Fourier transform of $D_1^{-1}$. However, this assumption holds for the case which I am most interested in, where $F$ is the matrix sign function: $(i\eta + \text{sign}(\gamma_5D_1))^{-1} = (-i\eta + \text{sign}(\gamma_5D_1))/(1+\eta^2)$. It is clear that, again as long as $D_2$ has the correct behaviour at small $p$, $\FT{\hat{B}}$ is analytic and thus  $\hat{B}$ is local for this particular range of values of $F$. Therefore, for this to be a valid renormalisation group transformation requires $|i\eta\gamma_5 + \gamma_5F(\gamma_5x_n (p)d_n(p))|>1\forall p$ for the non-zero modes\footnote{For conciseness, I use a potentially confusing shorthand: by non-zero modes of $D_2$ I mean all eigenvectors/eigenvalues of $D_2$ excluding the zero modes and their partners at eigenvalue 2.} in the limit that $\eta \rightarrow 0$. This means, in particular, that $F$ cannot cross zero at any point. However, I have already stated that $F$ must be an odd function. Therefore, $F(x)$ must be discontinuous at $x=0$.

Now I consider the specific case that $F(x) = \sign(x)$. Using the formulation of appendix \ref{app:B}, where the Dirac operator is written in the basis of the eigenvector pairs, I can write, for the non-zero eigenvectors,
\begin{align}
i\eta\gamma_5 &+ \gamma_5 F(\gamma_5 D_1) = \frac{1}{4}\left[\frac{\lambda^2}{2} - 1 + i\gamma_5 \left(\eta + g_2\lambda\sqrt{1-\frac{\lambda^2}{4}}\right)\right].\label{eq:63}
\end{align} 
The eigenvalues of this matrix are
\begin{gather}
\mu = \frac{\lambda^2}{2} - 1 \pm i \sqrt{\eta^2 + \lambda^2\left(1-\frac{\lambda^2}{4}\right)},
\end{gather}
which gives $|\mu|^2 = 1+\eta^2 $, which is greater than one.
For the zero modes and their partners at eigenvalue 2, I obtain\footnote{Here and in subsequent sections I assume that the zero modes, $|\psi_0\rangle$, have a positive chirality, i.e. $\gamma_5 |\psi_0\rangle =  |\psi_0\rangle$; the case when they have negative chirality can easily be considered using the same method and will give the same result.}
\begin{gather}
i\eta\gamma_5 + \gamma_5 F(\gamma_5 D_1) =  - \gamma_5 +i\eta\gamma_5 ,\label{eq:65}
\end{gather}
which gives eigenvalues
$|\mu|^2 = 1+\eta^2 > 1$. Therefore all the eigenvalues of $i\eta\gamma_5 + \gamma_5 F(\gamma_5 D_1)$ are larger than 1 and the inverse blocking is local. 

\subsection{Positivity of $\hat{B}\hat{\overline{B}}$}\label{sec:positivity}
The eigenvalue spectrum of $\hat{B}\hat{\overline{B}}$ is identical to the eigenvalue spectrum of $B'=(D_2+i\eta\gamma_5)^{-1/2}Z^{\dagger}(D_0 + i\eta\gamma_5)Z(D_2 + i \eta\gamma_5)^{-1/2}$. Thus proving that $\hat{B}\hat{\overline{B}}$ is positive is equivalent to proving that $B'$ is positive.
If $B'$ is positive, then
\begin{gather}
\psi^{\dagger}(x)(B')[x,x'] \psi(x') > 0\label{eq:whatever}
\end{gather}
for every possible non-zero $\psi$. I define
\begin{gather}
\psi_n = \int d^4x\theta\left(\frac{1}{2} - |x-n|\right) \psi(x),
\end{gather}
and write
\begin{gather}
e_n^{\psi} = \int d^4 x \psi_n(x)  B_x(x,n).
\end{gather}
The calculation of equation (\ref{eq:whatever}) proceeds in precisely the same way as for the Fourier transforms of $\hat{B}$ and $\hat{\overline{B}}$: in fact, baring the replacement of $e_n(p)$ with $e_n^{\psi}$ the integrals are precisely those needed for the Fourier transforms, but at $p=0$. Thus
\begin{gather}
\psi^{\dagger}(x)(B')[x,x'] \psi(x') = \sum_n
\frac{e_n^{\psi} (e_n^{\psi})^{\dagger}}{e_n(p=0) (e_n(p=0))^{\dagger}}\FT{B'}(p=0).
\end{gather}
Since, as already established, $D_2$ has the correct continuum limit, $\FT{B'}(p=0)\sim1$ (any difference from 1 will be due to a fermion renormalisation constant, which will be positive). Furthermore, $[e_n^{\psi} (e_n^{\psi})^{\dagger}] > 0$ and $[e_n (e_n)^{\dagger}]>0$ unless either $e_n^{\psi} = 0$, which is impossible for every $n$ for non-zero $\psi$ unless $D_2$ contains a zero mode with no equivalent in $D_0$ (if $\sum_n [e_n^{\psi} (e_n^{\psi})^{\dagger}] = 0 \forall n$ then $\psi^{\dagger}D_2\psi = 0$, so this condition would mean that there was an additional zero mode of $D_2$ generated by the lattice artefacts).

The proof that $B \overline{B}$ is positive proceeds in the same way. 
This completes the proof that this is a valid renormalisation group transformation as long as the lattice spacing is sufficiently fine that the indexes of $D_0$ and $D_2$ are equal.
\subsection{The Ginsparg-Wilson symmetry}\label{sec:5.5}
Following the notation of appendix \ref{app:B}, I decompose the continuum Dirac operator into eigenvector pairs and zero modes, and, in the basis of one of the eigenvector pairs, it can be written as 
\begin{gather}
D_0 +i\eta\gamma_5= \left[ \gamma_5 g_2 \lambda  + i\eta\gamma_5\right].
\end{gather}
$D_2$ is defined as
\begin{gather}
D_2 = 1 +  i\eta\gamma_5  + \gamma_5 \text {sign}\;(\gamma_5 (D_1 - B_C)).
\end{gather}
The non-zero pairs of $D_2$ can be decomposed according to equations (\ref{eq:63}) and (\ref{eq:65}).
\begin{gather}
D_2+i\eta\gamma_5 = \left[\frac{\lambda^2}{2} + \gamma_5 g_2 \left(\lambda\sqrt{1 - \frac{\lambda^2}{4}}\right)\right]+i\eta\gamma_5\label{eq:matrix_form_of_D2}.
\end{gather}
For the zero modes and their partners at eigenvalue 2, I can write
\begin{gather}
D_2 = 1+\gamma_5 + i\gamma_5\eta.\label{eq:matrix_form_of_D2b}
\end{gather}
From equation (\ref{eq:19}), the terms entering the Ginsparg-Wilson equation were given as $\overline{S}=\overline{B}\Gamma_5\hat{\overline{B}}$ and $S=\hat{B}\Gamma_5 B$. In a trivial topological sector, 
\begin{align}
S =& (D_2+i\gamma_5 \eta)^{-1}(D_0+i\eta\gamma_5)\gamma_5 (D_0+i\eta\gamma_5)^{-1}  (D_2+i\gamma_5\eta) \nonumber\\
=& - (D_2+i\gamma_5\eta)^{-1}\gamma_5 (D_2+  i\gamma_5 \eta)+ O(\eta)\nonumber\\
=& \gamma_5(1 - D_2) + O(\eta),\nonumber\\
\overline{S}=&\gamma_5,\label{eq:S}
\end{align}
where the first  equality follows from $\{D_0,\gamma_5\} = 0$, and the second follows from equation (\ref{eq:matrix_form_of_D2}). These blockings, of course, lead to the familiar form of the Ginsparg-Wilson equation. 

\subsection{Non-trivial Topology}
In the presence of zero modes, the above analysis breaks down because unless the zero modes of $D_0$ map precisely to the zero modes of $D_2$, the blocking matrix will contain a singularity. If $| \phi_0\rangle$ is a zero mode of $D_0$ and not a zero mode of $D_2$ (which will be the case unless the lattice spacing is zero) then clearly $\langle \phi_0| (D_0+i\eta\gamma_5)^{-1} (D_2+i\eta\gamma_5)$ is proportional to $1/\eta$. Therefore, it is necessary to introduce an additional term, the $Z$ of equation (\ref{eq:overlapblocking1}), to make $|\phi_0\rangle$ finite, either by mapping the zero modes of $D_0$ to the zero modes of $D_2$, by multiplying the zero modes by $\eta$ or some combination of the two. This $Z$ clearly has to be local, and leave the properties of the blocking described in the previous sections unaffected. 

I define the non-zero eigenvalue pairs of $D_0$ as $\frac{1}{\sqrt{2}}\left(|\phi_+^i\rangle \pm |\phi_-^i\rangle\right)$, and the eigenvectors of $D_2$ as $\frac{1}{\sqrt{2}}\left(|\psi_+^i\rangle \pm |\psi_-^i\rangle\right)$, $|\psi_0\rangle$ and $|\psi_2\rangle$, where the last two vectors refer to the zero mode and its partner. These vectors are defined so that $\gamma_5|\psi_{\pm}\rangle = \pm |\psi_{\pm}\rangle$. I have assumed for simplicity in this notation that the topological index is one, although the argument can be extended for all possible topological indices. 

The purpose of the unitary operator $Z$ is to render $\langle\phi_0|(D_0 + i\eta\gamma_5)^{-1} Z (D_2+i\eta\gamma_5)$ and $\langle{\psi_0}|(D_2+i\eta\gamma_5)^{-1}Z^{\dagger} (D_0+i\eta\gamma_5)$ finite as $\eta\rightarrow0$. The simplest construction which achieves this is
\begin{gather}
\hat{Z} = |\phi_0\rangle\langle\psi_0| + g_2|\phi_0\rangle\langle\psi_2|+ |\phi_{+}^i\rangle u^{ij}_+\langle\psi_+^j| + |\phi_{-}^i\rangle u^{ij}_-\langle\psi_-^j|, 
\end{gather}
where $u_{\pm}$ is some pair of unitary matrices. This maps the zero modes of $D_0$ onto the zero modes of $D_2$, thus ensuring that the blocking is valid. However, it is not clear that this example is local, and I have been unable to prove (or disprove) its locality. Instead, I shall use the construction 
\begin{align}
Z = &Z_3\frac{1}{\sqrt{Z_3^{\dagger} Z_3}}\nonumber\\
Z_3 =& \frac{1}{2}\left[\text{sign}(\gamma_5 D_0 - i\eta)\text{sign}(\gamma_5 D_3 - i \eta) +\text{sign}(\gamma_5 D_0 + i\eta)\text{sign}(\gamma_5 D_3 + i \eta)\right]\nonumber\\
=&|\phi_+^i\rangle\langle\phi_-^i|\psi_-^j\rangle\langle\psi_+^j| + |\phi_-^i\rangle\langle\phi_+^i|\psi_+^j\rangle\langle\psi_-^j| -|\phi_0\rangle\langle\phi_0|\psi_0\rangle\langle\psi_0| - \frac{\eta}{\lambda_2^j\sqrt{1-(\lambda_2^j)^2/4}}|\phi_0\rangle\langle\phi_0|\psi_+^j\rangle\langle\psi_+^j|- \nonumber\\
&\phantom{space}\frac{\eta}{\lambda_0^i}|\phi_+^i\rangle\langle\phi_+^i|\psi_0\rangle\langle\psi_0| -  \frac{\eta}{\lambda_0^i}|\phi_-^i\rangle\langle\phi_-^i|\psi_2\rangle\langle\psi_2|
,\label{eq:Z}
\end{align}
where $D_3 = D_2 - D_2^{\dagger}$ and I have assumed that $\gamma_5|\psi_0\rangle = |\psi_0\rangle$. $\eta$ is again an infinitesimal parameter, and I have neglected terms of O($\eta^2$). A similar expression can be easily constructed for the opposite chiral sector, and it will give the same final results.
 I define the matrix sign function of the non-Hermitian operator as $\text{sign}(A) = A / \sqrt{A^{\dagger} A}$. The matrix sign function is known to be exponentially local for $A^{\dagger}A\neq0$, and 
\begin{align}
Z_3^{\dagger}Z_3 =\frac{1}{4}&\left[ 2 + \text{sign}(\gamma_5 D_0 + i\eta)\text{sign}(\gamma_5D_3 + i\eta)\text{sign}(\gamma_5D_3 - i\eta)\text{sign}(\gamma_5D_0 - i\eta) +\right.\nonumber\\& \left.\text{sign}(\gamma_5 D_0 - i\eta)\text{sign}(\gamma_5D_3 - i\eta)\text{sign}(\gamma_5D_3 + i\eta)\text{sign}(\gamma_5D_0 + i\eta)\right],
\end{align}
which is manifestly greater than zero, which means that $Z_3/\sqrt{Z_3^{\dagger}Z_3}$ is analytic. Therefore this $Z$ is local. It is tedious but trivial to demonstrate using the locality of $Z$ and the technology of the previous section that the conditions required for a valid blocking hold. Note that $[\gamma_5,Z_3]=[\gamma_5,Z]=0$. This is, of course, not the only possible $Z$ which can be used to generate overlap fermions, nor is it likely to be the best. It is simply offered as an illustration that functions with the desired properties exist.

The continuum chiral symmetry is given by equation (\ref{eq:chiral_symmetry}). Therefore the operators required for the Ginsparg-Wilson equation are
\begin{align}
\overline{S} =& Z^{\dagger}\gamma_5(1-|\phi_0\rangle\langle\phi_0|)Z = \gamma_5(1-|\psi_0\rangle\langle\psi_0|) + O(\eta)\nonumber\\
S=& D_2^{-1}Z^{\dagger}D_0\gamma_5(1-|\phi_0\rangle\langle\phi_0|)D_0^{-1} Z D_2\nonumber\\
=&\gamma_5(1-D_2)(1-|\psi_0\rangle\langle\psi_0| - |\psi_2\rangle\langle\psi_2|) - \gamma_5|\psi_2\rangle\langle\psi_2| +O(\eta).
\end{align}
These operators will be local, despite their apparent dependence on the eigenvectors of $D_2$, because the blockings are local.
Therefore the Ginsparg-Wilson equation reads 
\begin{align}
0=\gamma_5(1-|\psi_0\rangle\langle\psi_0|) D_2 + D_2 \gamma_5 \left[(1-D_2)(1-|\psi_0\rangle\langle\psi_0|- |\psi_2\rangle\langle\psi_2|) - |\psi_2\rangle\langle\psi_2| \right].
\end{align}
This is equivalent to the standard form of the Ginsparg-Wilson relation, although this construction leads to slightly non-standard projection operators. However the chiral symmetry transformation operators  are different, and explicitly depend on the zero modes and their partner.  The associated Ginsparg-Wilson chiral symmetry is given by
\begin{align}
\overline{\psi} \rightarrow& \overline{\psi}e^{i\upsilon\gamma_5(1+|\psi_0\rangle\langle\psi_0)};& \psi \rightarrow& e^{i\upsilon[\gamma_5(1-D)(1-|\psi_0\rangle\langle\psi_0|- |\psi_2\rangle\langle\psi_2|) + \gamma_5 (|\psi_0\rangle\langle\psi_0|- |\psi_2\rangle\langle\psi_2|])}\psi.
\end{align}
\subsection{Summary}
The main conclusion of this work so far is that overlap fermions satisfy the Ginsparg-Wilson relation. This can, of course, also be established by simpler methods.
 
I have demonstrated that equation (\ref{eq:overlapblocking1}) describes a valid renormalisation group blocking for the case when $F(x) =\sign(x)$. I have also constructed the Ginsparg-Wilson equation associated with this blocking, and written down the corresponding symmetry and topological charge. The Dirac operator generated by this blocking is, in the limits that $\zeta \rightarrow 0$ and $\eta \rightarrow 0$, the familiar form of the massless overlap operator. This can be seen by noting that $\sign(\gamma_5 (D_1 - B_C)) = \sign(\gamma_5 D_1) - \gamma_5 B_C + O(e^{-\eta})$, and $1-B_C$ and $\sign(\gamma_5 D_1)$ are exponentially suppressed away from the lattice sites. Since, in this limit, the Dirac operator is proportional to  $\delta(x - na)\delta(y - nb)$, the continuum action $\int d^4x d^4y \overline{\psi}_2 (x)D_2(x,y) \psi_2(y)$ reduces to the lattice action, $\sum_{n,n'}\overline{\psi}_2 (n)D_2(n,n') \psi_2(n)$. 

As $\zeta \rightarrow\infty$ this operator, $D_2$, is the lattice overlap operator. It is zero everywhere except at the lattice sites, and it has precisely the same form when connecting fermion fields on lattice sites as Neuberger's original operator. Thus $D_2^{-1}$ is ill defined away from the lattice sites in the $\zeta\rightarrow\infty$ limit. However, the Fourier transform of $D_2^{-1}$ remains analytic, except for the usual pole at $p=0$. This apparently paradoxical conclusion depends on two properties of the (twisted mass regulated) matrix sign function: firstly that its inverse can be constructed from the Wilson operator without any contribution from the inverse Wilson operator, and secondly that its eigenvalues are always greater than one. This means that the Taylor series expansion which is needed for the evaluation of the Fourier transform is in ($(1- B_C)\sign(\gamma_5 D_1)$ rather than $\sign(\gamma_5 D_1)-B_C$, where $B_C$, which becomes $-1$ off the lattice sites, is the term which gives the non-lattice site positions their value of zero. Because the series expansion is now $B_C\sign(\gamma_5 D_1)$ rather than $B_C$ or $B_C F(D_W^{-1})$ it converges rather than diverges at $B_C=1$, allowing the Fourier transform to remain finite. 

\section{The symmetric blocking}\label{sec:7}
I noted earlier that for each Dirac operator there are an infinite number of blockings which could be used to generate that operator. Each of these blockings will lead to a different Ginsparg-Wilson relation for the same Dirac operator, and thus, as noted recently by Mandula~\cite{Mandula:2007jt,Mandula:2009yd}, there is actually an infinite group of lattice chiral symmetries; each with its own bare current (although from this work it is clear that, since the Ginsparg-Wilson relations are related by various renormalisation group transformations, the renormalised currents in a fixed renormalisation scheme must be identical). This degeneracy follows naturally from from the renormalisation group construction of the symmetry (see equation \ref{eq:degenerateblockings}), and it is not just present in lattice gauge theories, but also continuum theories as soon as I go beyond an ultra-local (i.e. point-like in the continuum) action. Thus this degeneracy must be present in any lattice gauge theory linked to the continuum via a renormalisation group blocking. 

The degeneracy, of course, arises because I have treated $\psi$ and $\overline{\psi}$ as independent variables, which is permitted in Euclidean, but not in Minkowski space-time. When constructing a chiral gauge theory, it is advantageous if the two Ginsparg-Wilson functions, $S$ and $\overline{S}$, are in some respect symmetric, so that, under $\mathcal{CP}$, $\overline{S}$ transforms into an operator proportional to $S$ and vice versa. I will now therefore consider the case of a symmetric blocking as a step towards constructing a chiral gauge theory.

In the trivial topological sector, the situation is reasonably straightforward. Consider the following blocking:
\begin{align}
B=&D_0^{-1/2} Z D_2^{1/2},\nonumber\\
\overline{B} =&  D_2^{1/2} Z^{\dagger} D_0^{-1/2},
\end{align}
where again, for this initial stage of the calculation, I shall use $Z=1$ before generalising. Proof that this blocking is valid proceeds using the same methods of section \ref{sec:5}. To construct the Ginsparg-Wilson equation and the chiral symmetry, I use the matrix decomposition of appendix \ref{app:B}. In my $\gamma$-matrix representation, I can write
\begin{align}
D_0^{-1/2} =& \frac{1}{\sqrt{2\lambda}}\left(\begin{array}{l l}
1&-1\\
1&1\end{array}
\right),\nonumber\\
D_2^{1/2} = &\sqrt{\frac{\lambda_2}{2}}\left(\begin{array}{l l}
\sqrt{1+\lambda_2/2}&\sqrt{1-\lambda_2/2}\\
-\sqrt{1-\lambda_2/2}&\sqrt{1+\lambda_2/2}\end{array}
\right).
\label{eq:matrixrep}
\end{align}

The Ginsparg-Wilson relation for this blocking can be derived from equations (\ref{eq:19}) and  (\ref{eq:matrixrep}). In particular, if there are no zero modes,    
\begin{align}
g_2=&Z^{\dagger}D_0^{-1/2}\gamma_5 D_0^{1/2}Z,\label{eq:88}\\
\overline{S} =& D_2^{\frac{1}{2}}Z^{\dagger} D_0^{-\frac{1}{2}} \gamma_5 D_0^{\frac{1}{2}}Z D_2^{-\frac{1}{2}} = 
\left(D_2 \frac{1}{\sqrt{D_2^{\dagger}D_2}} \right)g_2,\nonumber\\
S = & D_2^{-\frac{1}{2}} Z^{\dagger}D_0^{\frac{1}{2}} \gamma_5 D_0^{-\frac{1}{2}}Z D_2^{\frac{1}{2}} =
-g_2\left(D_2 \frac{1}{\sqrt{D_2^{\dagger}D_2}} \right).
\label{eq:Shalf}
\end{align} 
In matrix notation,
\begin{align}
-g_2D_2 \frac{1}{\sqrt{D_2^{\dagger}D_2}} =& \left(\begin{array}{l l}
\sqrt{1-\frac{\lambda^2}{4}}&-\frac{\lambda}{2}\\
-\frac{\lambda}{2}&-\sqrt{1-\frac{\lambda^2}{4}}
\end{array}\right),\nonumber\\
D_2 \frac{1}{\sqrt{D_2^{\dagger}D_2}} g_2 = &\left(\begin{array}{l l}
\sqrt{1-\frac{\lambda^2}{4}}&\frac{\lambda}{2}\\
\frac{\lambda}{2}&-\sqrt{1-\frac{\lambda^2}{4}}
\end{array}\right).\label{eq:Smatrix}
\end{align}
The Ginsparg-Wilson relation is
\begin{gather}
0=\left(D_2 \frac{1}{\sqrt{D_2^{\dagger}D_2}}\right)g_2 D_2 - D_2g_2\left( D_2 \frac{1}{\sqrt{D_2^{\dagger}D_2}}\right). \label{eq:symmetricGW}
\end{gather}
I note that in the limit of vanishing lattice spacing (where $a\lambda \rightarrow 0$), both $\overline{S}$ and $S$ reduce to $\gamma_5$, and the Ginsparg-Wilson relation reduces to the standard continuum chiral symmetry. 

Once again, working in a non-trivial topological sector means re-introducing the $Z$ operator of section \ref{sec:5}, equation (\ref{eq:Z}). This modifies equation (\ref{eq:Shalf}) to 
\begin{align}
\overline{S} =&   
\left(1-|\psi_0\rangle\langle\psi_0| - |\psi_2\rangle\langle\psi_2|\right)\left(D_2 \frac{1}{\sqrt{D_2^{\dagger}D_2}} \right)g_2 - \sqrt{\frac{i\eta}{2}}|\psi_0\rangle\langle\psi_2| - \sqrt{\frac{2}{i\eta}}|\psi_2\rangle\langle\psi_0| ,\nonumber\\
S = &
-g_2\left(D_2 \frac{1}{\sqrt{D_2^{\dagger}D_2}} \right)\left(1-|\psi_0\rangle\langle\psi_0| - |\psi_2\rangle\langle\psi_2|\right)+ \sqrt{\frac{2}{i\eta}}|\psi_0\rangle\langle\psi_2| + \sqrt{\frac{i\eta}{2}}|\psi_2\rangle\langle\psi_0|.
\label{eq:Shalftopnontrivial}
\end{align}
This formulation, though it follows from the renormalisation group derivation, is not obviously local, since it depends on the zero modes and their partners which are, in general, non-local. The relationship between this operator and $\gamma_5$ is also unclear. To make the relationship between these renormalisation group operators and $\gamma_5$ clearer, it is possible to re-write equation (\ref{eq:Shalf}), which in this construction only applies for the non-zero modes, in the form
\begin{align}
\overline{S} = & \gamma_5 \sqrt{1-\frac{D_2^{\dagger} D_2}{4}} + \frac{1}{4}\gamma_5(D_2 - D_2^{\dagger})\frac{1}{\sqrt{1-\frac{D_2^{\dagger} D_2}{4}}}\nonumber\\
S = &\gamma_5 \sqrt{1-\frac{D_2^{\dagger} D_2}{4}} - \frac{1}{4}\gamma_5(D_2 - D_2^{\dagger})\frac{1}{\sqrt{1-\frac{D_2^{\dagger} D_2}{4}}},\label{eq:S2}
\end{align}
and then extend this definition to the non-trivial topological sector.\footnote{I hypothesise that there exists some $Z$ which maps $S$ and $\overline{S}$ to these forms, but I have not yet found an explicit form for it. For this reason I also present the more cumbersome, but known to be valid, form of equation (\ref{eq:Shalftopnontrivial}).}  It is manifest that these operators are Hermitian; furthermore it can be shown using the standard form of the Ginsparg-Wilson relation that $S^2 = \overline{S}^2 = 1$. There is a question over the locality of these operators. If the term inside the square roots is positive, then, following an argument similar to the square root in the definition of overlap fermions, we can expect the operators to be local. The eigenvalues of $D_2^{\dagger}D$ are constrained between $0$ and $4$\footnote{There is a question concerning the mass regularisation, which increases the eigenvalues of $D_2^{\dagger} D_2$ by $\eta^2$. However, equation \ref{eq:S2} was explicitly calculated at $\eta = 0$. The equivalent expression for finite $\eta$  contains the square root of $1-\frac{D_2^{\dagger} D_2 - \eta^2}{4}$. Therefore the mass regularisation does not affect the discussion on locality one way or the other.}. The only difficulty is for the partners of the zero modes, where both $D_2 - D_2^{\dagger}$ and $\sqrt{1-D_2^{\dagger}D/4}$ are zero. We know that $\gamma_5(D_2 - D_2^{\dagger})/{\sqrt{1-{D_2^{\dagger} D_2}/{4}}}$ is well defined for the eigenvalues of $D_2^{\dagger} D_2$ at 4 because $S^2 = 1$, however this does not demonstrate that this operator, which is constructed from the matrix sign function of a shifted overlap operator, is local. However, given that numerical results have shown that the overlap operator is local even when one of its kernel eigenvalues is zero, I can expect that a similar result will hold for this operator. This is an issue which must be investigated further. 

The Ginsparg-Wilson relation is given by
\begin{align}
\overline{S} D_2 + D_2 S = 0,\label{eq:GW5}
\end{align}
and again a short calculation shows that any Dirac operator obeying the standard Ginsparg-Wilson relation also obeys this equation.

The chiral symmetry transformations are
\begin{align}
\overline{\psi} \rightarrow &\overline{\psi}e^{-i\upsilon\overline{S}}\nonumber\\
\psi\rightarrow & e^{i\upsilon S}  \psi.
\end{align} 
 In appendix \ref{app:CP4} I demonstrate that these blockings transforms under $\mathcal{CP}$ according to
\begin{align}
\mathcal{CP}[S] =& -W^{-1}\overline{S}^T W,\nonumber\\
\mathcal{CP}[\overline{S}] =& -W^{-1}{S}^T W.\label{eq:gamma5hatcp}
\end{align}
In the continuum, of course, $S = \overline{S} = \gamma_5$, and the same transformation properties hold. This suggests that it might be possible to construct a chiral gauge theory from these blockings, and this is the topic of the following sections. 

Leaving the considerations concerning the zero-modes aside, because $B$ and $\overline{B}$ are local and invertible, $\overline{S}$ and $S$ are local and invertible, and thus the projection operators in the chiral gauge theory will be local.
\section{Chiral gauge theory}\label{sec:8}
Previous discussions of lattice chiral gauge theory and renormalisation group blockings can be found in~\cite{Hasenfratz:2007dp,Gattringer:2007dz}.

In the continuum, the chiral gauge theory Lagrangian is\footnote{The notation used in this section is described in appendix \ref{app:CP}.} 
\begin{gather}
\mathcal{L}_0 =\frac{1}{4} \overline{\psi}_0(1+\gamma_5)D_0(1-\gamma_5)\psi_0 + \frac{1}{4}\overline{\psi}_0(1-\gamma_5)D_0(1+\gamma_5)\psi_0. 
\end{gather}
Once again, I can apply the renormalisation group blocking, $\psi_0 = B\psi_2$ and $\overline{\psi}_0 = \overline{\psi}_2\overline{B}$, to obtain the new Lagrangian
\begin{gather}
\mathcal{L}_2= \frac{1}{4}\overline{\psi}_2(1+\overline{B}\gamma_5\hat{\overline{B}})D_2(1-\hat{B}\gamma_5B)\psi_2 + \frac{1}{4}\overline{\psi}_2(1-\overline{B}\gamma_5\hat{\overline{B}})D_2(1+\hat{B}\gamma_5B)\psi_2.
\end{gather} 
One can therefore write lattice projectors
\begin{align}
P_{\pm} = & \frac{1}{2}(1\pm S)\nonumber\\
\overline{P}_{\pm} = & \frac{1}{2}(1\pm \overline{S})\label{eq:97}.
\end{align} 

It is a well-known problem that with the standard formulation of the Ginsparg-Wilson chiral symmetry (discussed in section \ref{sec:5}), the chiral formulation of the action violates $\mathcal{CP}$ symmetry~\cite{Hasenfratz:2001bz,Fujikawa:2002is,Fujikawa:2002vj}. With the standard projector operators for the overlap operator, it is clear why this is the case. 
The chiral Lagrangian is
\begin{gather}
\mathcal{L}_- = \overline{\psi}\; \overline{P}_+ D_2 P_- \psi.\label{eq:chiralaction}
\end{gather}
Using the traditional form of the  Ginsparg-Wilson symmetry, the projectors are
\begin{align}
P_{\pm} =& \frac{1}{2}(1\pm\gamma_5)\nonumber\\
\overline{P}_{\pm} = &\frac{1}{2}(1\pm \gamma_5(1-D_2)),
\end{align}
which means that the transformation of the Lagrangian under $\mathcal{CP}$ is
\begin{gather}
\mathcal{CP}(\mathcal{L}_-) = \overline{\psi} \gamma_5P_+\gamma_5 D_2 \overline{P}_-\psi,
\end{gather}
and it immediately follows that this action is not invariant under $\mathcal{CP}$ because of the anti-symmetry between $P$ and $\overline{P}$. It has been shown that, with the canonical form of the Ginsparg-Wilson equation, a chiral gauge theory satisfying $\mathcal{CP}$ cannot be constructed~\cite{Fujikawa:2002is}. 
Modifying the Projectors so that they are symmetric, $P_{\pm} = \overline{P}_{\pm} = (1\pm\gamma_5(1-D/2))/2$ fails because $|\gamma_5(1-D/2)|\neq 1$, and indeed can be zero, which leads to a non-locality. However, I do not use the canonical Ginsparg-Wilson equation, but the modified form of  equation (\ref{eq:GW5}), and this allows me to avoid the cited no-go theorem for chiral gauge theories.

If the Ginsparg-Wilson equation derived in the previous section is used (equation (\ref{eq:S2})), different projectors will result, 
and the chiral projectors can be constructed easily according to equation (\ref{eq:97}).
From the matrix representation of equation (\ref{eq:Smatrix}), it is clear that $(S)^2 = 1$, $(\overline{S})^2 = 1$, $\gamma_5 S \gamma_5 = \overline{S}$, $(S)^{\dagger} = S$ and  $(\overline{S})^{\dagger} = \overline{S}$, and this can also be proved directly using the Ginsparg-Wilson relation. Thus suitable projectors can be formed from these operators.  I use the projectors
\begin{align}
\overline{P}_{\pm} =& \frac{1}{2}\left(1\pm\left(\gamma_5 \sqrt{1-\frac{D_2^{\dagger} D_2}{4}} + \frac{1}{4}\gamma_5(D_2 - D_2^{\dagger})\frac{1}{\sqrt{1-\frac{D_2^{\dagger} D_2}{4}}}  \right)\right)\nonumber\\
P_{\pm} =& \frac{1}{2}\left(1\pm \left( \gamma_5 \sqrt{1-\frac{D_2^{\dagger} D_2}{4}} - \frac{1}{4}\gamma_5(D_2 - D_2^{\dagger})\frac{1}{\sqrt{1-\frac{D_2^{\dagger} D_2}{4}}} \right)\right)
\end{align}
The chiral Lagrangian can be written as
\begin{gather}
\overline{\psi} D \psi = \overline{\psi}\;\overline{P}_+D P_-\psi + \overline{\psi}\;\overline{P}_-D P_+\psi 
\end{gather}
and the equality can be demonstrated using the Ginsparg-Wilson relation. From equation (\ref{eq:gamma5hatcp}), $\mathcal{CP}[P_+] = W^{-1}\overline{P}_-^T W$ and $\mathcal{CP}[P_-] = W^{-1}\overline{P}_+^T W$, so each term in this Lagrangian transforms correctly under $\mathcal{CP}$. Therefore, this is, potentially, a suitable chiral gauge theory Lagrangian. 

\section{The non-Abelian gauge anomaly}\label{sec:8b}
In this section, which is intended as no more than a preliminary exploration of the topic, I shall only consider the topological trivial sector, leaving other sectors for subsequent work. I shall also now switch to the lattice theory so that the measure can be well defined non-perturbatively. In~\cite{Luscher:1999un}, Martin L\"uscher discussed Weyl fermions on the lattice and the non-Abelian gauge anomaly. The issue is that the new Weyl fermion fields given by $\Psi=P_-\psi$, $\overline{\Psi} = \overline{\psi}\;\overline{P}_+$ will, in general, have a measure which depends on the gauge field because the projectors depend on the gauge field. Thus after a gauge transformation (for example), the measure is not obviously invariant, which may give rise to an anomaly, which would have to be canceled in the variation in the fermion determinant.

We can select basis vectors $v_i$ and $\overline{v}_j$ such that
\begin{align}
v_i =& P_-v_i, & \overline{v}_j =& \overline{P}_+ \overline{v}_j,\\
(v_i,v_j) =&\delta_{ij}& (\overline{v}_i,\overline{v}_j) =& \delta_{ij}\label{eq:defnOfv}
\end{align}
The fermion fields $\Psi$ and $\overline{\Psi}$ can then be written as
\begin{align}
\Psi =& c_i v_i& \overline{\Psi} = & \overline{c}_j \overline{v}_j
\end{align}
for some coefficients $c$ and $\overline{c}$. The measure will then be $dc\;d\overline{c}$. If we pass to a different basis, $v_i \rightarrow v'_i = v_j Q^{-1}_{ji}$ (with a similar change in basis for $\overline{v}$), the measure will change by 
\begin{gather}
\delta_Q \mathbb{L}_{\upsilon} = \ln \det [Q] - \ln\det[\overline{Q}],\label{eq:detQ}
\end{gather}
with $\mathbb{L}_{\upsilon}$ defined below in equation (\ref{eq:Leta}), and, since $Q$ and $\overline{Q}$ are unitary, this will be a pure phase.  The expectation value of an observable $O$ can be given as
\begin{gather}
\langle O\rangle = \frac{1}{Z} \int D[U] e^{-S_g}\langle O\rangle_f.
\end{gather}
In a trivial topological sector, the fermionic expectation value is
\begin{gather}
\langle O\rangle_f = \int d\Psi d\overline{\Psi} O e^{-S_f},
\end{gather}
where $S_g$ is the gauge action and $S_f$ the fermionic action. For example, the fermionic propagator is given by
\begin{gather}
\langle\overline{\Psi}(x)\Psi(y)\rangle_f = \langle 1\rangle_f P_- S(x,y) P_+,
\end{gather}
where $S$ is the Green's function associated with the Dirac operator $D$, and
\begin{align}
\langle 1\rangle_f =& \det M, & M_{kj} =& \overline{v}_k D v_j.\label{eq:effectiveaction}
\end{align}
If we consider infinitesimal variations of the gauge field, such as
\begin{gather}
\delta_{\upsilon}U(x,\mu) = \upsilon^a_{\mu}T^a (x)U(x,\mu), 
\end{gather}
where $T^a$ are the (anti-Hermitian traceless) generators of the gauge group in some suitable representation $R$, then the variation of the effective action, defined in equation (\ref{eq:effectiveaction}) will be
\begin{gather}
\delta_{\upsilon}\ln\det M = \Tr[(\delta_{\upsilon} D) P_ D^{-1} \overline{P}_+] - i \mathbb{L}_{\upsilon},\label{eq:luscher4.4}
\end{gather}
where 
\begin{gather}
\mathbb{L}_{\upsilon} = i (v_j,\delta_{\upsilon} v_j) - i (\delta_{\upsilon}\overline{v}_j,\overline{v}_j).\label{eq:Leta}
\end{gather}
The first term in equation (\ref{eq:luscher4.4}) is obtained from the variation of $M$, the second from the variation of the measure. A current can be defined from $\mathbb{L}_{\upsilon}$, 
\begin{gather}
\mathbb{L}_{\upsilon} = \upsilon^a_{\mu}(x) j^a_{\mu}(x)
\end{gather}
and L\"uscher's first requirement for a valid chiral gauge theory is that this current is local. His second was that the measure should respect the gauge covariance, which means that equation (\ref{eq:luscher4.4}) should be zero when $\upsilon$ is a gauge transform
\begin{align}
\upsilon_{\mu}(x) =& - \nabla_{\mu} \omega(x)\nonumber\\
\nabla_{\mu} \omega(x)=&U_{\mu}(x)\omega(x+\hat{\mu})U^{\dagger}_{\mu}(x) - \omega(x).\label{eq:111}
\end{align}
L\"uscher's third condition considers paths in the space of possible gauge fields. If we write the gauge field as $U^t$, where $t$ indicates the location on a smooth curve in configuration space, then for a closed loop (running from $t = 0$ to $t = 1$), we can define a Wilson line
\begin{align}
W = &e^{i\int_0^1 dt \mathbb{L}_{\upsilon}},& \upsilon_{\mu}(x) =& \partial_t (U^t_{\mu}(x)) (U^t_{\mu}(x))^{-1}.\label{eq:Luscher6.2}
\end{align}
Writing $\overline{P}^t$ and $\hat{P}^t$ as the projectors associated with a gauge field $U^t$, and defining the unitary operators,
\begin{align}
\partial_t\hat{Q}^t =&[\partial_t \hat{P}^t,\hat{P}^t]\hat{Q}^t &\hat{Q}^0 = &1,\nonumber\\
\partial_t\overline{Q}^t =&\overline{Q}^t[\partial_t \overline{P}^t,\overline{P}^t] &\overline{Q}^0 = &1,\label{eq:Luscher6.4}
\end{align}
from which it can be proved that
\begin{align}
(\hat{Q}^t)^{-1} P_t Q^t = & \hat{P}^0 \nonumber\\
\overline{Q}^t P_t (\overline{Q}^t)^{-1} = &\overline{P}^0.\label{eq:Luscher6.5}
\end{align}
L\"uscher's third condition is that, for a closed loop, $W$ should be independent of the path used to travel from $U^0$ to $U^t$.  

In this section, I intend to begin a discussion of how my new construction fits into this framework, although I will here limit myself to a discussion of the current and variation of the measure under gauge transformations, neglecting the subsequent elements of the original discussion. However, it is necessary to show that the three conditions are satisfied.

First of all, I need to choose the basis vectors $v$ and $\overline{v}$, and it is particularly convenient to construct this basis from the eigenvector pairs of $H=\gamma_5 D_2$. In my matrix notation, I rewrite the Dirac operator as
\begin{gather}
H = \lambda \left(\begin{array}{c c}
\frac{\lambda}{2}&\sqrt{1-\frac{\lambda^2}{4}}\\
\sqrt{1-\frac{\lambda^2}{4}}& -\frac{\lambda}{2} 
\end{array}\right).
\end{gather}
For simplicity, I re-write this in terms of an angle $\theta$, where $\cos\theta = \lambda/2$ and $\sin\theta = \sqrt{1-\lambda^2/4}$. Then,
\begin{align}
\overline{S} = & \left(\begin{array}{c c}
\sin\theta & \cos\theta\\
\cos\theta & - \sin\theta
\end{array}\right),\nonumber\\
S = & \left(\begin{array}{c c}
\sin\theta & -\cos\theta\\
-\cos\theta & - \sin\theta
\end{array}\right)
\end{align}
and the eigenvectors of $S$ and $\overline{S}$ are (suppressing the eigenvector index)
\begin{align}
\left(\begin{array}{c} |\overline{S}_+\rangle\\
|\overline{S}_-\rangle\end{array}\right) = & 
\left(\begin{array}{c c}
\cos(\pi/4 - \theta)&-\sin(\pi/4 - \theta)\\
\sin(\pi/4 - \theta)&\cos(\pi/4 - \theta)
\end{array}\right)
 \left(\begin{array}{c} |H_+\rangle\\
|H_-\rangle\end{array}\right),\nonumber\\
\left(\begin{array}{c} |{S}_+\rangle\\
|{S}_-\rangle\end{array}\right) = & 
\left(\begin{array}{c c}
\cos(\pi/4)&\sin(\pi/4)\\
-\sin(\pi/4)&\cos(\pi/4)
\end{array}\right)
 \left(\begin{array}{c} |H_+\rangle\\
|H_-\rangle\end{array}\right),
\end{align}
where $|H_+\rangle$ and $H_-\rangle$ are the eigenvectors of $H$ with positive and negative eigenvalue respectively, and $|S_{\pm}\rangle$ and $|\overline{S}_{\pm}\rangle$ are similarly the positive and negative eigenvectors of $S$ and $\overline{S}$. We can then choose the basis such that $\overline{v} = |\overline{S}_+\rangle$ and $v = |S_-\rangle$. Differentiating the eigenvectors using the procedure outlined in~\cite{Cundy:2007df}, gives
\begin{gather}
\delta_{\upsilon}|H_\pm\rangle = (1-|H_{\pm}\rangle\langle H_{\pm}|) \frac{1}{H \mp \lambda} \delta H |H_{\pm}\rangle,
\end{gather}   
and 
\begin{align}
(\delta_{\upsilon}\overline{v},\overline{v}) =& \sin\left(\frac{\pi}{4} - \theta\right)\cos\left(\frac{\pi}{4} - \theta\right)\left(\langle H_+|\frac{1}{2\lambda}\delta_{\upsilon} H |H_-\rangle - \langle H_-|\frac{1}{2\lambda}\delta_{\upsilon} H |H_+\rangle\right)\nonumber\\
({v},\delta_{\upsilon}{v}) =& -\sin\left(\frac{\pi}{4}\right)\cos\left(\frac{\pi}{4}\right)\left(\langle H_+|\frac{1}{2\lambda}\delta_{\upsilon} H |H_-\rangle - \langle H_-|\frac{1}{2\lambda}\delta_{\upsilon} H |H_+\rangle\right).
\end{align}
Therefore,
\begin{align}
\mathbb{L}_{\upsilon} =& i \frac{\lambda}{8}\left(\langle H_-|\delta_{\upsilon} H |H_+\rangle - \langle H_+|\delta_{\upsilon} H |H_-\rangle\right)\nonumber\\
=&\frac{i}{8} \Tr[\delta_{\upsilon} (D_2) (\overline{S} - S)].
\end{align}
Using equations (\ref{eq:GW5}) and (\ref{eq:luscher4.4}) and the result $S D_2^{\dagger} = (D_2-D_2^{\dagger})\gamma_5/2$, it is possible to show that
\begin{gather}
\delta_{\upsilon}\ln\det M = \frac{1}{2}\Tr\left[\delta_{\upsilon}(D_2) D_2^{-1}\right] + \frac{1}{4}\Tr\left[\delta_{\upsilon}(D_2)\left(\frac{1}{D}\overline{S} - S\frac{1}{D_2}\right)\right]+\frac{1}{8} \Tr[\delta_{\upsilon} (D_2) (\overline{S} - S)] ,\label{eq:condition1}
\end{gather}
and the current $j_{\mu}(x)$ is defined as
\begin{gather}
\Tr[\upsilon_{\mu}(x) j_{\mu}(x)] = \frac{1}{8}\Tr[\delta_{\upsilon} (D_2) (\overline{S} -S)].\label{eq:condition2}
\end{gather}
I need to demonstrate that $j_{\mu}$ is local and $\delta_{\upsilon}\ln\det M = 0$ when $\upsilon$ represents a gauge transform. The proof of the second condition is straight-forward. The first term in equation (\ref{eq:condition1}) is $\delta_{\upsilon}\Tr[\ln D_2]$, and, since the eigenvalues of $D_2$ are invariant under a gauge transformation, this is clearly zero.  For the second and third terms in equation (\ref{eq:condition2}), I use the result~\cite{Luscher:1999un} that for a gauge transformation,
\begin{gather}
\delta_{\upsilon}(D_2) = [ R(\omega),D_2],
\end{gather}
where $R$ is the representation of the SU(3) generators, and the infinitesimal gauge transformation $\upsilon$ is defined in terms of $\omega$ through equation (\ref{eq:111}).
Using the Ginsparg-Wilson equation, I now write the second term in equation (\ref{eq:condition1}) as 
\begin{align}
\frac{1}{4}\Tr\left[\delta_{\upsilon}(D_2)\left(\frac{1}{D_2}\overline{S} - S\frac{1}{D_2}\right)\right] =& \frac{1}{4}\Tr [R(\omega)(\overline{S} + S)]\nonumber\\
=&\frac{1}{2}\Tr\left[R(\omega)\gamma_5\sqrt{1-\frac{D^{\dagger}D}{4}}\right] = 0
\end{align}
in the topological trivial sector.
The third term is proportional to
\begin{gather}
\mathbb{L}_{\upsilon} = -\frac{i}{8}\Tr\left[R(\omega)\gamma_5 D^{\dagger} D \sqrt{1-\frac{D^{\dagger}D}{4}}\right] = 0,
\end{gather}
because $\Tr \gamma_5 = 0$ and everything else within the trace commutes with $\gamma_5$.
Thus both the effective action and the measure of the chiral gauge theory are gauge-invariant.

We can ask how the measure changes under a change in the gauge field. Equations (\ref{eq:Luscher6.4}) and (\ref{eq:Luscher6.5}) describes the unitary operators $\hat{Q}$ and $\overline{Q}$ which evolve the projectors under a change of the gauge field. Using equation (\ref{eq:detQ}), the change in $\mathbb{L}_{\upsilon}$ is given by $\ln \det{\hat{Q}}  - \ln\det{\overline{Q}}$. Since all gauge transformations within a topological sector are connected, it suffices to consider an infinitesimal change in the gauge field from a `time' $t$ to a `time' $t + \delta t$. Then, from equation (\ref{eq:Luscher6.4}), and implicitly summing over the eigenvector indices $i$ and $j$,
\begin{align}
\delta (Q_t) =& (1-2 \hat{P}_t) \delta(P_t) Q_t\nonumber\\
=&(|S_+^i\rangle\langle S_+^i| - |S^i_-\rangle\langle S^i_-|)(\delta(|S^j_-\rangle)\langle S^j_-| + |S^j_-\rangle \delta(\langle S^j_-|)).
\end{align}
Hence,
\begin{align}
\det(Q_t + \delta Q) =& \det[Q]\det[|S_+^i\rangle\langle S_+^i|  + |S^i_-\rangle\langle S^i_-| +\nonumber\\
&\phantom{det[Q]\det[S]} |S_+^i\rangle\langle S_+^i|(\delta(|S^j_-\rangle)\langle S^j_-| -  |S^i_-\rangle(\langle S_-^i|\delta(|S^j_-\rangle)\langle S^j_-| + \delta (\langle S^i_-|))]\nonumber\\
=& \det[Q] (1 -  \langle S_-^i|\delta(|S^i_-\rangle) - \delta (\langle S^i_-|) | S^i_-\rangle).
\end{align}
Since $\langle S_-^i|\delta(|S^i_-\rangle)=0$, $\det[Q_{t + \delta t}] = \det[Q_t]$, and by induction, as $Q_0$ is defined as 1, $\det Q_t = 1$ for all $t$. This means that $\mathbb{L}_{\upsilon}$ remains zero when the gauge field is evolved, and the Wilson loop from equation (\ref{eq:Luscher6.2}) is always 1. It obviously follows that the value of the Wilson loop  is independent of the path, and L\"uscher's third condition for a valid chiral gauge theory is satisfied.

I now need to demonstrate that the current associated with the gauge transformation is local. I use the integral representation of the matrix sign function,
\begin{gather}
\sign(\xi) = \frac{\xi}{\pi}\int_{-\infty}^{\infty} dt (t^2 + \xi^2)^{-1}, 
\end{gather}
and the current associated with the transformation of the gauge fields is generated by
\begin{align}
4\pi\kappa \Tr[j_{\mu}\upsilon_{\mu}] =& \delta_{\upsilon}(H_1)\left( t\frac{1}{t^2 + H_1^2}(\overline{S} - S) \frac{1}{t^2 + H_1^2} t -  H_1\frac{1}{t^2 + H_1^2}(\overline{S} - S) \frac{1}{t^2 + H_1^2} H_1\right),\label{eq:current}
\end{align}
where $H_1$, given explicitly below in equation \ref{eq:H1}, satisfies $H_1= \gamma_5 D^{Lattice}_1(n,n')$, where $D^{Lattice}$ corresponds to the operator between lattice sites $n$ and $n'$. Using the notation of equation (\ref{eq:Ireallydontwanttocalculatethis}), $D^{Lattice}_1(n,n') = (\overline{B}_x)^{-1}D_1B_x^{-1}$
Once again, I require a slightly redefined form of the Wilson blockings, this time making them functions of $t$ by modifying the mass term:
\begin{align}
B_W(y,x)(t) = &\sum_n \zeta^4 e^{-\zeta \sum_{\mu}|x_{\mu} - n_{\mu}|}
\prod_{\beta,\gamma}\theta\left(\frac{1}{2} - |x_{\gamma} - n_{\gamma}|\right)
\theta\left(\frac{1}{2} - |y_{\beta} - n_{\beta}|\right)
\nonumber\\
&e^{-\gamma_{\mu}(m+it\gamma_5)(1+r^2 + 3\epsilon r^2)(y_{\mu}-n_{\mu})/(1+2r^2\epsilon)}\left(1 + r \sum_{\theta} \gamma_{\theta}N(y_{\theta} - n_{\theta})\right)\nonumber\\ &
\sum_{\mathfrak{L}_{x,n},\mathfrak{L}_{n,y}}e^{-W[\mathfrak{L}_{x,n}]}U[\mathfrak{L}_{xn}]e^{-W[\mathfrak{L}_{n,y}]}U[\mathfrak{L}_{n,y}] ,\nonumber
\\
\overline{B}_W(x',y)(t) = &\sum_n \zeta^4 e^{-\zeta \sum_{\mu}|x_{\mu} - n_{\mu}|}
\prod_{\beta,\gamma}\theta\left(\frac{1}{2} - |x_{\gamma} - n_{\gamma}|\right)\theta\left(\frac{1}{2} - |y_{\beta} - n_{\beta}|\right)\nonumber\\
&\left(1 - r \sum_{\theta} \gamma_{\theta}N(y_{\theta} - n_{\theta})\right)e^{\gamma_{\mu}(m+it\gamma_5)(1+r^2 + 3\epsilon r^2)(y_{\mu}-n_{\mu})/(1+2r^2\epsilon)}\nonumber\\
&\sum_{\mathfrak{L}_{x,n},\mathfrak{L}_{n,y}}e^{-W[\mathfrak{L}_{x,n}]}U[\mathfrak{L}_{xn}]e^{-W[\mathfrak{L}_{n,y}]}U[\mathfrak{L}_{n,y}].
\end{align}
The inverse Wilson blockings can be constructed as in equation (\ref{eq:inverseBlockings}). Then, defining $H_1$ as the lattice part of the Hermitian Wilson operator,
\begin{align}
H_{1}(n,n') =& \gamma_5 \int d^4y \theta\left(\frac{1}{2} - |y_{\beta} - n_{\beta}|\right)\left(1 - r \sum_{\theta} \gamma_{\theta}N(y_{\theta} - n_{\theta})\right)\nonumber\\
&e^{\gamma_{\mu}(m+it\gamma_5)(1+r^2 + 3\epsilon r^2)(y_{\mu}-n_{\mu})/(1+2r^2\epsilon)}e^{-W[\mathfrak{L}_{n,y}]}U[\mathfrak{L}_{n,y}]D_0\bigg[\theta\left(\frac{1}{2} - |y_{\beta'} - n'_{\beta'}|\right)\nonumber\\
&e^{-\gamma_{\mu}(m+it\gamma_5)(1+r^2 + 3\epsilon r^2)(y_{\mu}-n'_{\mu})/(1+2r^2\epsilon)}\left(1 + r \sum_{\theta} \gamma_{\theta}N(y_{\theta} - n'_{\theta})\right) e^{-W[\mathfrak{L}_{n',y}]}U[\mathfrak{L}_{n',y}]\bigg],\label{eq:H1}
\end{align}
I can write that
\begin{align}
\big[\gamma_5\overline{B}_W&(t)D_0B_W(t)\gamma_5\overline{B}_W(-t)D_0B_W(-t)\big](x,x') =& \overline{B}_x(x,n)(H_{1}^2 + t^2)B_x(x',n') x_n,
\end{align}
where $B_x$ and $x_n$  are defined in and above equation (\ref{eq:Ireallydontwanttocalculatethis}) together with $d_n$ which I shall require shortly. 
Then, if $S_0$ is the Greens function associated with $D_0$,
\begin{gather}
(H_{1}^2 + t^2)^{-1} = x_n^{-1} B_x B_W^{-1}(-t)S_0 \overline{B}_W^{-1}(-t)\gamma_5B_W^{-1}(t)S_0\overline{B}_W^{-1}(t)\gamma_5 \overline{B}_x.\label{eq:monster3}
\end{gather}
Finally, to construct the current itself, I assume that $\upsilon$ is invertible, and write that
\begin{gather}
\delta_{\upsilon}D_1 = \upsilon_{\mu} \Delta_{\upsilon}\left[D_1\right],\label{eq:monster1},
\end{gather}
where 
\begin{gather}
\Delta_{\upsilon} = \lim_{\upsilon \rightarrow 0}(\upsilon_{\mu})^{-1}(e^{\int d^4\chi \upsilon_{\mu}(\chi)\frac{\partial\phantom{A_{\mu}(\chi)}}{\partial A_{\mu}(\chi)}} - 1),
\end{gather}
and I use
\begin{gather}
\overline{S} - S = \frac{1}{2}(D_2 - D_2^{\dagger})\gamma_5\left(1- \frac{D_2^{\dagger}D_2}{4}\right)^{-1/2}.\label{eq:monster2}
\end{gather}
It is now merely a matter of straight-forward algebra to determine whether the current is local. I simply Taylor expand the square root in equation (\ref{eq:monster2}), and using either a polynomial or the integral representation of the matrix sign function in $\overline{S}-S$, insert equations (\ref{eq:monster2}) and (\ref{eq:monster1}) into equation (\ref{eq:current}) to extract the current, then Fourier transform the current using the methods of section \ref{sec:5}, recombine the polynomial or integrate over the dummy variables to recreate the sign functions, and finally test to see whether the result is analytic.  
Writing
\begin{align}
s_n(t) &= \int d^4y e^{-ipn'} \theta\left(\frac{1}{2} - |y_{\beta} - n_{\beta}|\right)\left(1 - r \sum_{\theta} \gamma_{\theta}N(y_{\theta} - n_{\theta})\right)\nonumber\\
&e^{\gamma_{\mu}(m+it\gamma_5)(1+r^2 + 3\epsilon r^2)(y_{\mu}-n_{\mu})/(1+2r^2\epsilon)}e^{-W[\mathfrak{L}_{n,y}]}U[\mathfrak{L}_{n,y}]S_0\bigg[\theta\left(\frac{1}{2} - |y_{\beta'} - n'_{\beta'}|\right)\nonumber\\
&e^{-\gamma_{\mu}(m+it\gamma_5)(1+r^2 + 3\epsilon r^2)(y_{\mu}-n'_{\mu})/(1+2r^2\epsilon)}\left(1 + r \sum_{\theta} \gamma_{\theta}N(y_{\theta} - n'_{\theta})\right) e^{-W[\mathfrak{L}_{n',y}]}U[\mathfrak{L}_{n',y}]\bigg] e^{i p n'}\gamma_5\nonumber\\
\intertext{and}
h_n(t)&=\gamma_5\int d^4y e^{-ipn'} \theta\left(\frac{1}{2} - |y_{\beta} - n_{\beta}|\right)\left(1 - r \sum_{\theta} \gamma_{\theta}N(y_{\theta} - n_{\theta})\right)\nonumber\\
&e^{\gamma_{\mu}(m + i t \gamma_5 )(1+r^2 + 3\epsilon r^2)(y_{\mu}-n_{\mu})/(1+2r^2\epsilon)}e^{-W[\mathfrak{L}_{n,y}]}U[\mathfrak{L}_{n,y}]D_0\bigg[\theta\left(\frac{1}{2} - |y_{\beta'} - n'_{\beta'}|\right)\nonumber\\
&e^{-\gamma_{\mu}(m + i t \gamma_5 )(1+r^2 + 3\epsilon r^2)(y_{\mu}-n'_{\mu})/(1+2r^2\epsilon)}\left(1 + r \sum_{\theta} \gamma_{\theta}N(y_{\theta} - n'_{\theta})\right) e^{-W[\mathfrak{L}_{n',y}]}U[\mathfrak{L}_{n',y}]\bigg] e^{i p n'},
\end{align}
the Fourier transform of the current is
\begin{align}
4\pi\kappa& \Tr [\FT{j_{\mu} \upsilon_{\mu}}] =  \int_{-\infty}^{\infty} dt  \delta_{\upsilon}(h_n(0))\bigg( t s_n(t) s_n(-t) (\gamma_5 \sign(\gamma_5x_nd_n) - \sign(\gamma_5x_nd_n)\gamma_5 )\nonumber\\
&\phantom{spacespace}\left(2 - \gamma_5 \sign(\gamma_5x_nd_n) - \sign(\gamma_5x_nd_n)\gamma_5  \right)^{-1/2}s_n(t)s_n(-t) t -\phantom{a}\nonumber\\
&h_n(0)s_n(t)s_n(-t) (\gamma_5 \sign(\gamma_5x_nd_n) - \sign(\gamma_5x_nd_n)\gamma_5 )\nonumber\\
&\phantom{spacespace}\left(2 - \gamma_5 \sign(\gamma_5x_nd_n) - \sign(\gamma_5x_nd_n)\gamma_5  \right)^{-1/2} s_n(t)s_n(-t) h_n(0)\bigg).\label{eq:upthere}
\end{align}
$\gamma_5$-Hermiticity and $\mathcal{CP}$ invariance of $\gamma_5 h_n(0)$ mean that the Fourier transform of $h_n(0)$ must be of the form $Z_0(p) =  - m\gamma_5 + \gamma_5Z_5(p)+ \gamma_5 \gamma_{\mu} Z_{\mu}(p)$, where $Z_{\mu}$ and $Z_5$ are real functions. Since $s_n$ is associated with the Green's function of $h_0$, it will have a simple pole at  $i t  +Z_0(p)$, and this will be the only potential non-analyticity in the current (given that the term which we take the inverse square root of is positive in this trivial topological sector and that the Wilson propagator only has one simple pole). $Z_0^2$ commutes with both $Z_0$ and $\gamma_5$. Thus we can write that $s_n(t) s_n(-t) = \varpi/(t^2 + Z_0^2)$, where $\varpi$ is some constant. Using the results,
\begin{align}
\int_{-\infty}^{\infty} dt \frac{t^2}{(t^2 + Z_0^2)^2} =& \frac{\pi}{2|Z_0|},\nonumber\\
\int_{-\infty}^{\infty} dt \frac{Z_0}{(t^2 + Z_0^2)^2} =& \frac{\pi}{2|Z_0|^2}.
\end{align}
I write that the Fourier transform of the current is proportional to
\begin{gather}
\frac{1}{((Z_5 - m)^2 + Z_{\mu}^2)^{2}} (\gamma_5\gamma_{\mu}Z_{\mu})(Z_5 - m)\gamma_5(\gamma_5\gamma_{\mu}Z_{\mu}),\nonumber
\end{gather}
where I have expanded $(2-\gamma_5 \sign (Z_0) - \sign(Z_0)\gamma_5)^{-1/2}$ around $Z_0 = 0$ and neglected the higher order contributions. This is analytic for real $p$\footnote{Except possibly on those gauge field configurations where the kernel operator has a precisely zero eigenvalue, a situation which has zero measure in the functional integration over gauge fields. This corresponds to the boundary between different topological sectors.}, and since there are no possible poles except at $Z_0 = 0$, this means that the current is local.  Thus this formalism satisfies L\"uscher's three criteria for a valid lattice chiral gauge theory. 
    
\section{Conclusion}\label{sec:9}
I have shown that, in Euclidean space-time, certain lattice Dirac operators are connected to the continuum Dirac operator by a renormalisation group transformation, and for overlap fermions (and, by extension, it is easy to show that other Ginsparg-Wilson fermions constructed from the overlap operator~\cite{Cundy:2008cs} can be constructed using similar blockings~\cite{Cundy:2008cn}) this transformation remains valid as I take the `lattice limit,' which it does not for other lattice fermions, as long as the lattice topological index matches the continuum topological charge. Using this method, I have proposed a formalism for a lattice chiral gauge theory, which obeys $\mathcal{CP}$ symmetry and which reduces to the continuum chiral gauge theory, at least in the trivial topological sector.

I have made a number of key assumptions, which need to be tested in specific circumstances: including that the topological charge of the overlap operator reduces to that of the continuum, and that both the matrix sign function of the Wilson operator and the overlap operator remain local even if the operator has a precise zero mode. I have only considered the chiral gauge theory in the sector with zero topological index. If this work is to be given credence, these assumptions and limitations need to be addressed in subsequent work.

This work is primarily intended as a theoretical study: to tie up a number of loose ends with the overlap formalism; and as such I have not considered whether there are any practical benefits to this work. Whether this method can be used to, for example, calculate renormalisation constants for overlap fermions or match a lattice renormalisation scheme to a continuum scheme is a matter which I leave to a future discussion. 

It has been questioned whether this work implies that overlap fermions are classically perfect (in contradiction to numerical experience). There are two responses to this: Firstly, renormalisation will be required to compare between the results at two different lattice spacings, and the renormalisation constants will depend on the scale, and, secondly, this work assumes that an overlap construction of the gauge action is used, while (to my knowledge) no full scale simulations have been performed using an overlap gauge action. The discrepancy between the gauge and fermion actions will lead to lattice artefacts.

However, this work does suggest that it might be possible to move from a properly tuned lattice theory to the continuum theory by a renormalisation group transformation, avoiding a continuum extrapolation; the difficulty being that the blocking requires the continuum gauge fields and not just the lattice links. Whether such a method exists or could be used  practically is, of course, a subject for future research.

\section*{Acknowledgements}
I am grateful of support from the DFG For 465. I am grateful for useful discussions with Falk Bruckmann, Andreas Sch\"afer and Martin L\"uscher, and correspondence on earlier drafts from David Adams, Wolfgang Bietenholz, Artan Borici, Maarten Golterman, Peter Hasenfratz and Yigal Shamir.

\appendix
\section{Notation}\label{app:A}
I use the following $\gamma$ matrix notation:
\begin{gather}
\gamma_5 = \left(\begin{array}{r r}
1 &0\\
0&-1
\end{array}\right)\\
\gamma_4 = \left(\begin{array}{r r}
0 &1\\
1&0
\end{array}\right)\\
\gamma_i = \left(\begin{array}{r r}
0 &-i\sigma_i\\
i\sigma_i&0
\end{array}\right),
\end{gather}
where $i = 1,2,3$ and $\sigma_i$ are the Hermitian form of the Pauli matrices. The other representation of the $\gamma$-matrices, $g_{\mu}$, is defined below.
\section{Overlap eigenvalue decomposition}\label{app:B}
The overlap operator is
\begin{gather}
D_2 =1+\gamma_5\sign(\gamma_5D_1),
\end{gather}
and the squared Hermitian overlap operator,
\begin{gather}
D_2D_2^{\dagger} = 2 + \gamma_5 \sign(\gamma_5 D_1) + \sign(\gamma_5 D_1)\gamma_5,
\end{gather}
commutes with $\gamma_5$. This means that the non-zero eigenvalues of $D_2D_2^{\dagger}$ are degenerate, and $D_2^{\dagger}D_2$ can be written in a chiral basis
\begin{gather}
D_2D_2^{\dagger} = \left(\begin{array}{l l}
\lambda^2&0\\
0&\lambda^2
\end{array}\right).
\end{gather}
The degenerate eigenvectors of $D_2D_2^{\dagger}$ are $|\psi_+\rangle$ and $|\psi_-\rangle$, where $\gamma_5|\psi_{\pm}\rangle = \pm|\psi_{\pm}\rangle$. Thus 
\begin{align}
\langle\psi_+|D_2D_2^{\dagger} |\psi_+\rangle =& \lambda^2 = 2 + 2 \langle\psi_+|\sign(\gamma_5 D_W) |\psi_+\rangle\nonumber\\
\langle\psi_-|D_2D_2^{\dagger} |\psi_-\rangle =& \lambda^2 = 2 - 2 \langle\psi_-|\sign(\gamma_5D_W) |\psi_-\rangle.
\end{align}  
Since the matrix sign function is Hermitian and given that $[\sign(\gamma_5 D_1)]^2 = 1$, I can  write, in some suitable and non-standard $\gamma$-matrix representation where (excluding some, as yet undefined, contribution from the zero modes and their partners) $g_2^{(2)} = \sum_i |\psi_+^i\rangle\langle\psi_-^i| + |\psi_-^i\rangle\langle\psi_+^i|$\footnote{I write this as $g_2$ rather than $g_4$ because of the way in which it transforms under $\mathcal{CP}$, see section \ref{app:CP2}}: 
\begin{gather}
\sign (\gamma_5 D_W) =\left(\begin{array}{l l}
\frac{\lambda^2}{2}-1&\lambda\sqrt{1-\frac{\lambda^2}{4}}\\
\lambda\sqrt{1-\frac{\lambda^2}{4}}&1-\frac{\lambda^2}{2}
\end{array}
\right),
 \end{gather}
Now, given that both $\gamma_2$ and $g_2^{(2)}$ are invertible, I can write that
\begin{gather}
g_2^{(2)}= \tilde{U}_2 \gamma_2  \hat{U}_2,
\end{gather}
for some matrices $\tilde{U}_2$ and $\hat{U}_2$. That $\gamma_2$ and $g_2^{(2)}$ are Hermitian means that $\hat{U}_2 = \tilde{U}_2^{\dagger}$. $\gamma_2^2=1$ and $(g_2^{(2)})^2=1$ force $\tilde{U}_2$ to be unitary. I can use these unitary $\tilde{U}_2$ operators to construct the complete representation of $\gamma$-matrices, $g_{\mu}^{(2)}= \tilde{U}_2 \gamma_{\mu}  \tilde{U}_2^{\dagger}$. It is easy to demonstrate that these satisfy the same anti-commutation relations as the standard $\gamma$-matrices. Since the $g$-matrices are constructed from the eigenvectors, $g_5^{(2)} = \sum_i(|\psi_+^i\rangle\langle\psi_+^i| - |\psi_-^i\rangle\langle\psi_-^i|)$ plus some contribution from the zero modes and their partners, I can write $g_5^{(2)} = \gamma_5$. This means that $[\tilde{U},\gamma_5] = 0$, so that $\tilde{U}_2$ can be decomposed as
\begin{gather}
\tilde{U}_2 = \left(\begin{array}{l l }
U_2&0\\
0&U_2^{\dagger}
\end{array}\right),
\end{gather}
for unitary $U_2$. 

It follows that, 
\begin{align}
D_2 =& \left(\frac{1}{2}D^{\dagger}D  + \tilde{U}_2\gamma_5\gamma_2\tilde{U}_2^{\dagger}\sqrt{D^{\dagger}D\left(1-\frac{D^{\dagger}D}{4}\right)}\right) \label{eq:113}
\end{align}
The zero mode and its partner at 2 also form a pair of eigenvalues, although this time the eigenvalues do not have the same magnitude. In the basis of the zero mode and its partner, the Dirac operator can be written as $1\pm\gamma_5$, with the sign depending on the chirality of the zero mode. 

Equally, the continuum Dirac operator has non-zero eigenvectors in degenerate pairs. The Dirac operator anti-commutes with $\gamma_5$, is anti-Hermitian, traceless, which means that $D_0$ can be written in matrix form in the basis of a pair of non-zero modes
\begin{gather}
D_0 = \left(\begin{array}{l l}
0&\lambda U^2_0\\
-\lambda (U_0^2)^{\dagger}&0
\end{array}
\right).
\end{gather}  
For simplicity, in sections \ref{sec:5}, \ref{sec:7} and \ref{sec:8}, I have omitted the superscript indicating which $g$-matrix is used (depending on which Dirac operator). It should be clear which of the two different $g$-matrices is intended in the main text.

$g_2^{(2)}$ can be written as
\begin{align}
g_2^{(2)} = -\frac{1}{4}(1-\gamma_5)&D_2(1+\gamma_5)\frac{1}{\sqrt{D_2^{\dagger}D_2(1-D_2^{\dagger}D_2/4)}} +\phantom{a}\nonumber\\& \frac{1}{4}(1+\gamma_5)D_2(1-\gamma_5)\frac{1}{\sqrt{D_2^{\dagger}D_2(1-D_2^{\dagger}D_2/4)}}+|\psi_0\rangle\langle\psi_2| + |\psi_2\rangle\langle\psi_0|\label{eq:g4}
\end{align}
with $g_2^{(0)}$ defined in a similar way.

\section{$\mathcal{CP}$}\label{app:CP}
Charge conjugation is defined as
\begin{align}
\psi(x)\rightarrow& -C^{-1}\overline{\psi}^T(x),&\overline{\psi}(x)\rightarrow &\psi(x)^TC,\nonumber\\
U(x,\mu)\rightarrow & U(x,\mu)^*,
\end{align}
where `$T$' denotes the transpose and `$*$' the complex conjugate, and the charge conjugation matrix $C$ satisfies
\begin{align}
C^{\dagger}C =& 1,&C^T =& -C,&C\gamma_{\mu}C^{-1} = &-\gamma_{\mu}^T,&C\gamma_5 C^{-1} = \gamma_5^T.
\end{align}
The Dirac operator $D_0$ transforms as
\begin{gather}
D_0[U](x,y) \rightarrow C^{-1}D[U^*](x,y)^T C,
\end{gather}
and it is straightforward to show, by expanding the matrix sign function in a polynomial series, that $D_1$, $B_C$ and $D_2$ must transform in the same way.

The Parity operation is defined as
\begin{align}
\psi(x) \rightarrow& \gamma_4 \psi(\overline{x}),&\overline{\psi}(x)\rightarrow& \overline{\psi}(\overline{x})\gamma_4,\nonumber\\
U(x,\mu)\rightarrow &U^P(x,\mu)=\left\{\begin{array}{l l}
U^{\dagger}(\overline{x} - a \hat{\mu},\mu)&\mu = 1,2,3\\
U(\overline{x},\mu)&\mu = 4
\end{array}\right.,
\end{align}
where
\begin{gather}
\overline{x} = (-x_1,-x_2,-x_3,x_4).
\end{gather}
In this case,
\begin{gather}
D_0[U](x,y)\rightarrow \gamma_4D[U^P](\overline{x},\overline{y})\gamma_4.
\end{gather}

The $\mathcal{CP}$ transformation can be defined as
\begin{align}
\psi(x)\rightarrow&-W^{-1}\overline{\psi}^T(\overline{x}),&W^T=&W,&\overline{\psi}(x)\rightarrow&\overline{\psi}^T(\overline{x})W,\nonumber\\
U(x,\mu)\rightarrow&U^{CP}(x,\mu),
\end{align}
where
\begin{align}
W^{\dagger}W=&1&W\gamma_{\mu}W^{-1} =& \left\{\begin{array}{l l}
\gamma_{\mu}^{T}&\mu = 1,2,3\\
-\gamma_{\mu}^T&\mu = 4\end{array}\right.& W\gamma_5W^{-1} = -\gamma_5^T.\label{eq:gammacp}
\end{align}
Under this transformation,
\begin{gather}
D[U](x,y) = W^{-1} D[U^{CP}](\overline{x},\overline{y})^T W.\label{eq:121}
\end{gather}
The continuum action transforms under $\mathcal{CP}$ according to
\begin{align}
\overline{\psi}(x)D_0[U](x,y)\psi(y)\rightarrow& \psi(\overline{x})^T W W^{-1} D_0[U^{CP}](\overline{x},\overline{y})W W^{-1} W \overline{\psi}(\overline{y})^T\nonumber\\
=&\overline{\psi}(\overline{y})D_0[U](\overline{y},\overline{x})\psi(\overline{x}),
\end{align}
and thus this action is invariant under $\mathcal{CP}$. Similarly, for the chiral decomposition of the action
\begin{align}
\overline{\psi}(x)D_0[U](x,y)\psi(y) = \frac{1}{4}\overline{\psi}(x)&(1+\gamma_5)D_0[U](x,y)(1-\gamma_5)\psi(y) + \nonumber\\&\frac{1}{4}\overline{\psi}(x)(1-\gamma_5)D_0[U](x,y)(1+\gamma_5)\psi(y), 
\end{align}
both of the Weyl fermion actions are invariant under CP.
\subsection{Transformation of $D_2$ under $\mathcal{CP}$}\label{app:CP1b}
\noindent\textbf{Theorem \ref{app:CP1b}:} 
\begin{quote}
$D_2$ \textit{ transforms under }$\mathcal{CP}$\textit{ according to}
\begin{gather}
\mathcal{CP}[D_2(x,y,U)] = W^{-1}D_2(\overline{x},\overline{y},U^{CP})^T W,\label{eq:D2CP}
\end{gather} 
\textit{if }$F(\gamma_5 D_1)$\textit{ is an odd function.}
\end{quote}

\noindent \textbf{Proof:}\\

First of all, it is necessary to determine how $B_W$ and $\overline{B}_W$, which are defined in equation (\ref{eq:45}) transform under $\mathcal{CP}$. $N(y_{\theta} - n_{\theta})$ is an odd function, which means that $\mathcal{CP}[\gamma_{\theta} N(y_{\theta} - n_{\theta})] = \gamma_{\theta}^T N(\overline{y}_{\theta} - \overline{n}_{\theta})$. Similarly, $\mathcal{CP}[\gamma_{\mu}y_{\mu}] = \gamma_{\mu}^T\overline{y}_{\mu}$. Those elements of $B_W$ which are even in the coordinates are obviously invariant under $\mathcal{CP}$. This means that
\begin{align}
\mathcal{CP}[B_W(y,x,U)] = & W^{-1}\overline{B}_W(\overline{y},\overline{x},U^{CP})^T W,\nonumber\\
\mathcal{CP}[\overline{B}_W(y,x,U)] = & W^{-1}B_W(\overline{y},\overline{x},U^{CP})^T W.
\end{align}
Hence,
\begin{align}
\mathcal{CP}[D_1(y,x,U)] =& \mathcal{CP}[\overline{B}_W(y,y',U)D_0(y',x',U)B_W(x',x,U)]\nonumber\\ 
=& W^{-1}(\overline{B}_W(\overline{y},\overline{y}',U^{CP})D_0(\overline{y}',\overline{x}',U^{CP})B_W(\overline{x}',\overline{x},U^{CP}))^{T}\nonumber\\ =&
W^{-1}D_1(\overline{y},\overline{x},U^{CP})^T W.
\end{align}
It can also be shown that
\begin{gather}
\mathcal{CP}[B_C(y,x,U)] = B_C(\overline{y},\overline{x},U^{CP})^T .
\end{gather}
$D_2$ is defined as
\begin{align}
D_2 = &1+\frac{1}{2}\gamma_5(F[\gamma_5 \hat{D}_1] + F[-\gamma_5 \hat{D}_1]) + \frac{1}{2}\gamma_5(F[\gamma_5 \hat{D}_1] - F[-\gamma_5 \hat{D}_1])\nonumber\\=& 1 + \gamma_5 \sum_{n=0,2,4,\ldots} c_n(\gamma_5 \hat{D}_1)^n + \gamma_5 \sum_{n=1,3,5,\ldots} c_n(\gamma_5 \hat{D}_1)^n,
\end{align}
where $\hat{D}_1 = D_1-B_C$
This gives
\begin{align}
\mathcal{CP}[D_2&(x,y,U)] \nonumber\\=&
 W^{-1}\left(1 - \gamma_5^T \sum_{n=0,2,4\ldots} c_n(\gamma_5^T \hat{D}_1(U^{CP})^T)^n+ \gamma_5^T \sum_{n=1,3,5\ldots} c_n(\gamma_5^T \hat{D}_1(U^{CP})^T)^n\right) W\nonumber\\
=&W^{-1}\left(1 - \frac{1}{2}\gamma_5(F[\gamma_5\hat{D}_1(U^{CP})] + F[-\gamma_5\hat{D}_1(U^{CP})]) +\phantom{a}\right.
\nonumber\\
&\left. \phantom{spacespacespacespacespacespac}\frac{1}{2}\gamma_5(F[\gamma_5\hat{D}_1(U^{CP}]) - F[-\gamma_5\hat{D}_1(U^{CP})])\right)^T W .
\end{align}
And the result follows directly.
\subsection{Proof that an operator expansion of $\Tr\;\log(\hat{B}B)$ cannot contain $F_{\mu\nu}\tilde{F}_{\mu\nu}$}\label{app:CP1}
\noindent\textbf{Theorem \ref{app:CP1}:} 
\begin{quote}
\textit{If both }$D$\textit{ and }$D_2$\textit{ transform in the standard way under }$\mathcal{CP}$\textit{ (equations (\ref{eq:121}) and (\ref{eq:D2CP})) then the expansion of }$\Tr\;\log(\hat{B}B)$\textit{ cannot contain a term proportional to }$F_{\mu\nu}\tilde{F}_{\mu\nu}$.
\end{quote}

\noindent \textbf{Proof:}\\

I consider how $\hat{\overline{B}}$ and $\hat{B}$ transform when the symmetry is applied. I define
 \begin{align}
\mathcal{CP}[\hat{\overline{B}}] =& \hat{\overline{B}}_{CP},\nonumber\\
\mathcal{CP}[\hat{B}] =& \hat{B}_{CP}.
\end{align}
Then, using $D_2 = \hat{\overline{B}}D_0\hat{B}$,
\begin{align}
\mathcal{CP}[D_2] =& W^{-1}D_2[U^{CP}](\overline{x},\overline{y})^T W\nonumber\\
=&\hat{\overline{B}}_{CP}W^{-1}D_0[U^{CP}](\overline{x},\overline{y})^T W \hat{B}_{CP}.
\end{align}
Thus,
\begin{align}
\mathcal{CP}[\hat{\overline{B}}[U](x,y)] =&W^{-1}\hat{B}[U^{CP}](\overline{x},\overline{y})^T W,
\nonumber\\
\mathcal{CP}[\hat{B}[U](x,y)] =&W^{-1}\hat{\overline{B}}[U^{CP}](\overline{x},\overline{y})^T W,
\end{align}
and, by expanding $\log(\hat{\overline{B}}\hat{B})$ in a polynomial, 
\begin{gather}
\mathcal{CP}[\log(\hat{\overline{B}}\hat{B})] = W^{-1}\log[\hat{B}[U^{CP}](\overline{x},\overline{y})^T\hat{\overline{B}}[U^{CP}](\overline{x},\overline{y})^T]W.
\end{gather} 
Using the cyclicity of the trace, that the trace of the transpose of a matrix is equal to the trace and by suitably redefining variables it immediately follows that $\Tr\;\log(\hat{\overline{B}} \hat{B})$ is invariant under $\mathcal{CP}$. But, as is well known, $\tilde{F}_{\mu\nu}F_{\mu\nu}$ is anti-symmetric under $\mathcal{CP}$. Therefore this term, and any other terms which are not invariant under $\mathcal{CP}$ are forbidden in the expansion of $\Tr(\log \hat{\overline{B}}\hat{B})$. $F_{\mu\nu}^{2}$ is invariant under $\mathcal{CP}$ and therefore allowed.

\subsection{Transformation of $g_{\mu}^{(2)}$ under $\mathcal{CP}$}\label{app:CP2}
\noindent\textbf{Theorem \ref{app:CP2}:} 
\begin{quote}
\textit{The alternative }$\gamma$\textit{ matrices, }$g_{\mu}$\textit{ have the same transformation properties under }$\mathcal{CP}$\textit{ as the standard }$\gamma$\textit{-matrices, }$\gamma_{\mu}$.
\end{quote}

\noindent \textbf{Proof:}\\

The definition of $g_2$ is given by equation (\ref{eq:g4}), and, excluding the zero mode term, this is:
\begin{align}
g_2^{(2)} = -\frac{1}{4}(1-\gamma_5)D_2(1+\gamma_5)&\frac{1}{\sqrt{D_2^{\dagger}D_2(1-D_2^{\dagger}D_2/4)}} +\phantom{a}\nonumber\\& \frac{1}{4}(1+\gamma_5)D_2(1-\gamma_5)\frac{1}{\sqrt{D_2^{\dagger}D_2(1-D_2^{\dagger}D_2/4)}}
\end{align}
It is straightforward to show, using equations (\ref{eq:gammacp}) and (\ref{eq:121}) that under $\mathcal{CP}$, 
\begin{align}
W \mathcal{CP}[g_{4}^{(2)}(x,y,U)]W^{-1} = & W\bigg[
-\frac{1}{4}(1+\gamma_5^T)D_2^T(1-\gamma_5^T)\left(\frac{1}{\sqrt{D_2^{\dagger}D_2(1-D_2^{\dagger}D_2/4)}}\right)^T +\phantom{a}\nonumber\\
& \frac{1}{4}(1-\gamma_5^T)D_2^T(1+\gamma_5^T)\left(\frac{1}{\sqrt{D_2^{\dagger}D_2(1-D_2^{\dagger}D_2/4)}}\right)^T
\bigg]W^{-1}\nonumber\\
=&W(g_2^{(2)})^TW^{-1}\label{eq:gundercp}
\end{align}
where the last identity uses $[D_2,D_2^{\dagger}D_2] = [D_2^{\dagger}D_2,\gamma_5] = 0$.

Writing $g_{\mu} = \tilde{U}\gamma_{\mu}\tilde{U}^{\dagger}$, we can derive the transformation properties of the $\tilde{U}$ matrices from how $g_2$ transforms. From this, it can be shown that the $g_{\mu}$ matrices transform under $\mathcal{CP}$ in the same way as $\gamma_{\mu}$.

\subsection{Transformation of zero modes of $D_2$ under $\mathcal{CP}$}\label{app:CP3}
\noindent\textbf{Theorem \ref{app:CP3}:} 
\begin{quote}
\textit{Under }$\mathcal{CP}$\textit{, the zero modes and their partners transform according to}
\begin{align}
\mathcal{CP}[|\psi_0\rangle] =& W^{-1}|\psi_2\rangle,&\mathcal{CP}[|\psi_2\rangle] = &W^{-1}|\psi_0\rangle.\label{eq:zeromodesunderCPfinalresult}
\end{align}
\end{quote}

\noindent \textbf{Proof:}\\

In section \ref{sec:8}, I required the transformation properties of the zero modes of $D_2$ and their partners under $\mathcal{CP}$. These vectors, $|\psi_0\rangle$, and $|\psi_2\rangle$ satisfy (in a sector with negative topological index)
\begin{align}
 \text{sign} (\gamma_5D_1)|\psi_0\rangle =& |\psi_0\rangle,& \text{sign} (\gamma_5D_1)|\psi_2\rangle =& |\psi_2\rangle\nonumber\\
 \gamma_5|\psi_0\rangle =& |\psi_0\rangle,&  \gamma_5|\psi_2\rangle =& -
|\psi_2\rangle\label{eq:zeromodesunderCP1}
\end{align}
 These two eigenvectors (along with the other zero modes and their partners if the topological index is greater than 1) are the only spinor fields which are simultaneously eigenvectors of $\gamma_5$ and $\text{\sign}(\gamma_5 D_1)$, and, indeed, $\gamma_5\text{\sign}(\gamma_5 D_1)\gamma_5$. Suppose that under $\mathcal{CP}$, $|\psi_0\rangle$ transforms to $|\psi_0^{CP}\rangle^T$ and $|\psi_2\rangle$ to $|\psi_2^{CP}\rangle^T$. Then equation (\ref{eq:zeromodesunderCP1}) will transform to
\begin{align}
 W^{-1}\gamma_5[\text{sign} (\gamma_5D_1)]^T\gamma_5 W|\psi_0^{CP}\rangle^T =& |\psi_0^{CP}\rangle^T,\nonumber\\ W^{-1}\gamma_5[\text{sign} (\gamma_5D_1)]^T\gamma_5 W|\psi_2^{CP}\rangle^T =& |\psi_2^{CP}\rangle^T\nonumber\\
 -W^{-1}\gamma_5^T W|\psi_0^{CP}\rangle^T =& |\psi_0^{CP}\rangle^T,\nonumber\\ -W^{-1}\gamma_5W|\psi_2^{CP}\rangle^T =& -
|\psi_2^{CP}\rangle^T.\label{eq:zeromodesunderCP2}
\end{align}
It is now easy to deduce that
\begin{align}
\gamma_5\text{sign} (\gamma_5D_1)\gamma_5W|\psi_0^{CP}\rangle =& W|\psi_0^{CP}\rangle,\nonumber\\ \gamma_5\text{sign} (\gamma_5D_1)\gamma_5 W|\psi_2^{CP}\rangle =& W|\psi_2^{CP}\rangle\nonumber\\
 \gamma_5W|\psi_0^{CP}\rangle =& -W|\psi_0^{CP}\rangle,\nonumber\\  \gamma_5W|\psi_2^{CP}\rangle =& 
W|\psi_2^{CP}\rangle,\label{eq:zeromodesunderCP3}
\end{align}
and the result follows.
\subsection{Proof of equation (\ref{eq:gamma5hatcp})}\label{app:CP4}
\noindent\textbf{Theorem \ref{app:CP4}:} 
\begin{quote}
\textit{Under }$\mathcal{CP}$\textit{, the Ginsparg-Wilson functions} $S$\textit{ and} $\overline{S}$ \textit{transform according to}
\begin{align}
\mathcal{CP}[S] =& -W^{-1}\overline{S}^T W,\nonumber\\
\mathcal{CP}[\overline{S}] =& -W^{-1}{S}^T W.\label{eq:gamma5hatcpt}
\end{align}
\end{quote}

\noindent \textbf{Proof:}\\

In section \ref{sec:8}, equation (\ref{eq:S2}), I defined $S$ and $\overline{S}$ as
\begin{align}
\overline{S} = &\gamma_5 \sqrt{1-\frac{D_2^{\dagger} D_2}{4}} + \frac{1}{4}\gamma_5(D_2^{\dagger} - D_2)\frac{1}{\sqrt{1-\frac{D_2^{\dagger} D_2}{4}}}\nonumber\\
S = &\gamma_5 \sqrt{1-\frac{D_2^{\dagger} D_2}{4}} - \frac{1}{4}\gamma_5(D_2^{\dagger} - D_2)\frac{1}{\sqrt{1-\frac{D_2^{\dagger} D_2}{4}}}.
\end{align}
and I required how these transformed under $\mathcal{CP}$. Given that $D^{\dagger}D$ commutes with $\gamma_5$, it is easy to show that
\begin{align}
W^{-1}\mathcal{CP}\left[\gamma_5 \sqrt{1-\frac{D_2^{\dagger} D_2}{4}}\right] W =& - \left( \gamma_5 \sqrt{1-\frac{D_2^{\dagger} D_2}{4}} \right)^T\nonumber\\
W^{-1}\mathcal{CP}\left[\gamma_5(D_2^{\dagger} - D_2)\frac{1}{\sqrt{1-\frac{D_2^{\dagger} D_2}{4}}}\right] W =&  \left( \gamma_5(D_2^{\dagger} - D_2)\frac{1}{\sqrt{1-\frac{D_2^{\dagger} D_2}{4}}} \right)^T,
\end{align}
and equation (\ref{eq:gamma5hatcpt}) follows immediately.

That the alternative definition of $S$ and $\overline{S}$ given in equation (\ref{eq:Shalftopnontrivial}) obeys $\mathcal{CP}$ follows from the results of the previous appendices.

\bibliographystyle{elsarticle-num}
\bibliography{new_GW_operator}

\end{document}